\documentclass[aip,preprint]{revtex4-1}

\usepackage{graphicx}% Include figure files
\usepackage{dcolumn}% Align table columns on decimal point
\usepackage{bm}% bold math
\usepackage[utf8]{inputenc}
\usepackage[T1]{fontenc}
\usepackage{subcaption}
\usepackage{mathptmx}
\usepackage{etoolbox}

\makeatletter
\def\@email#1#2{%
 \endgroup
 \patchcmd{\titleblock@produce}
  {\frontmatter@RRAPformat}
  {\frontmatter@RRAPformat{\produce@RRAP{*#1\href{mailto:#2}{#2}}}\frontmatter@RRAPformat}
  {}{}
}%
\makeatother
\begin{document}

\preprint{AIP/123-QED}

\title[Runaway electron–whistler interactions]{Non-diffusive pitch angle scattering of runaway electrons in the presence of whistler waves}
% Force line breaks with \\
\author{Yashika Ghai}
 \email{ghaiy@ornl.gov}
\affiliation{ Oak Ridge National Laboratory, Oak Ridge, Tennessee 37831, USA}

\author{D. Del-Castillo-Negrete}
\affiliation{The University of Texas at Austin, Austin, Texas 78712, USA}

\author{D. A. Spong}
\affiliation{ Oak Ridge National Laboratory, Oak Ridge, Tennessee 37831, USA}

\author{M. T. Beidler}
\affiliation{ Oak Ridge National Laboratory, Oak Ridge, Tennessee 37831, USA}

\date{\today}% It is always \today, today,
             %  but any date may be explicitly specifie

\begin{abstract}

Wave-particle interactions between high-energy electrons and whistler waves lead to pitch-angle scattering of electrons in space, astrophysical, and fusion plasmas. In this work, we present a first-principles-informed computational framework to study runaway electron (RE)–whistler interactions in realistic 3D tokamak geometry, enabling investigation of RE pitch angle transport beyond the assumptions of quasilinear theory. This framework is realized by coupling AORSA, which solves the full-wave Maxwell–Vlasov system for whistler eigenmodes in a given tokamak equilibrium, with KORC, a kinetic orbit code that follows full-orbit RE trajectories in prescribed wave fields. Statistical analysis of the pitch-angle distribution moments and displacement variance is used to characterize the RE transport behavior. We find that REs undergo significant pitch-angle scattering and exhibit energy-dependent, non-diffusive transport behavior. In particular, we observe transitions between diffusive, sub-diffusive, and super-diffusive transport regimes depending on the initial RE kinetic energy, a behavior not captured within the standard quasilinear diffusion framework. These results provide new physical insights into runaway-electron transport driven by wave–particle interactions, and are relevant to the development of wave-based control schemes for runaway electrons in fusion plasmas.

\end{abstract}
\thanks{
Notice: This manuscript has been authored by UT-Battelle, LLC, under contract DE-AC05-00OR22725 with the US Department of Energy (DOE). The US government retains and the publisher, by accepting the article for publication, acknowledges that the US government retains a nonexclusive, paid-up, irrevocable, worldwide license to publish or reproduce the published form of this manuscript, or allow others to do so, for US government purposes. DOE will provide public access to these results of federally sponsored research in accordance with the DOE Public Access Plan (https://www.energy.gov/doe-public-access-plan).}
\maketitle

\section{Introduction}
Resonant wave-particle interactions---specifically between whistler waves and energetic electrons---are among the most fundamental mechanisms governing pitch angle scattering and energetic particle transport in space and astrophysical plasma environments \cite{space_Cattell, space_Gary_1977, space_Komarov_2018, space_res_Karimabadi,space_res_Kitamura2022, space_res_Li}.
Tokamak plasmas may undergo an abrupt loss of confinement, known as a 
disruption, during which the plasma rapidly loses its thermal energy in a phenomenon known as
thermal quench. The rapid decrease in plasma temperature increases the plasma 
resistivity and is followed by a rapid decay of the plasma current. The associated change in magnetic flux induces a large toroidal 
loop voltage and electric field. Electrons for which the electric-field 
acceleration exceeds the collisional drag can continuously gain energy and are 
referred to as runaway electrons (REs). REs can reach relativistic energies 
($1$--$20$~MeV) and may cause severe damage to plasma-facing components by 
depositing their energy on the device walls. Frontier science experiments performed in DIII-D \cite{spong}, for the first time, detected runaway electron-driven whistler waves in a tokamak, analogous to whistlers driven by lightning-induced high-energy electrons in near-Earth environments \onlinecite{armstrong}.  

\par
 DIII-D experiments demonstrated that REs undergo significant pitch-angle scattering through interactions with the destablized whistler waves, which can increase the perpendicular velocity of the RE population. Wave-induced pitch-angle scattering of low-energy REs ($\sim 2.5$ MeV) was also observed through interactions with lower-hybrid waves in the HT-7 tokamak~\cite{Chen_2010}. This effect is potentially important for RE control because increasing the perpendicular velocity enhances synchrotron emission power~\cite{Diego_synch_2018}, thereby increasing radiation damping—the only known intrinsic damping mechanism for runaway electron energy. These findings suggest that externally launched whistler waves may offer a viable mitigation strategy for disruption-generated REs in tokamak fusion devices, motivating the need for quantitative modeling of the underlying wave–particle interaction dynamics.
 
% The generation of runaway electrons following tokamak disruptions poses a significant threat to safe and reliable operation of future fusion reactors. High energy runaway beams that act like a welding torch can damage plasma-facing components and jeopardize reactor integrity, making effective mitigation strategies critical for realizing commercially viable fusion energy. To address this, the physics of avoidance and suppression of runaway electron beams has been widely studied by several authors in different tokamak devices \cite{vlasenkov1973,gill1993,plyusnin2006,riccardo2010disruption,arnoux2011heat}. Established mitigation techniques include resonant magnetic perturbations (RMPs)\cite{lehnen2008,koslowski2014,yoshino2000,gobbin2018} , massive gas injection (MGI) \cite{bozhenkov2008,pautasso2020,shevelev2021,commaux2011,lehnen2009runaway,reux2015} and shattered pellet injection \cite{commaux2016,li2018,Reux_2022}. 
 
Wave–particle interactions involving relativistic electron dynamics in both space~\cite{Horne,Thorne, Thorne_2021, Menietti_2021, Li_2014, Tao_zonca_2025} and fusion plasmas have been investigated, primarily within the framework of quasilinear diffusion models and simplified geometries. A linear stability analysis of the destabilization of magnetosonic-whistler waves by runaway electrons in fusion and astrophysical plasma environments was presented in Ref.~[\onlinecite{Tunde}]. The quasilinear treatment developed in that work showed that the dominant effect of the resulting whistler instability is enhanced pitch-angle scattering of electrons. In addition, the instability threshold for runaway-electron-driven waves, including whistlers and magnetized plasma waves, was analyzed using the COIN ray-tracing code~\cite{aleynikov}. More recently, a local instability analysis of RE-driven waves has been reported \cite{Castaldo_2024}, including electron and ion Bernstein waves (EBWs and IBWs), in the FTU tokamak plasma with a hot Maxwellian background, and identified whistler waves as the dominant instability driven by REs in the cold background plasma limit. 
 \par
Quasilinear diffusion models have also been used by several authors to analyze the role of kinetic whistler-wave instabilities in runaway electron avalanches. Simulations based on these models reported scattering of REs to large pitch angles in the presence of whistler fields and linked this effect to the high-frequency nonthermal ECE signals observed experimentally, which depend on both RE energy and pitch angle~\cite{changliu1,changliu2}. In addition, a mechanism for limiting runaway electron energies in post-disruption plasmas to a few MeV—thereby mitigating potential damage to plasma-facing components (PFCs)—has been explored through modifications of the RE momentum-space distribution induced by pitch-angle scattering in the presence of externally launched whistler fields~\cite{guo}.
 
 More recently, a marginal stability constraint in steady-state has been derived for the excitation of whistler and compressional Alfvén waves by the runaway electron distribution function \cite{breizman2023}. The authors employed a kinetic equation for runaway electrons with quasilinear diffusion and considered a balance of kinetic instability drive and collisional damping of the runaway distribution to derive the stability constraint.
%; under the assumption that synchrotron radiation is the main channel of particle energy loss and the radial transport of REs is negligible. 
In another study, a qualitative picture of runaway electrons dynamics in fusion reactor-relevant plasmas has been reported using particle-in-cell simulations in a 2-D, slab geometry \cite{Kang_2024}. This was based on evolving runaway electrons momentum distribution function at two initial kinetic energies (1 MeV and 10 MeV) in the presence of whistler fields, and it was concluded that the perpendicular diffusion of the RE momentum or pitch angle scattering is associated with the growth of whistler waves. 
  
% Allanson et al. \onlinecite{Allanson2019} performed particle-in-cell simulations and tracked the pitch angle diffusion of an ensemble of $10^8$ electrons in presence of an incoherent spectrum of whistler mode waves in a simple cold plasma with uniform background magnetic field. Their results showed that the nature of diffusive response in pitch-angle space is a function of charged particle energy and pitch angle and that nature of pitch angle transport is anomalous in some regions of the phase space. However, the model assumptions and parameters considered in the model were applicable in the magnetosphere regions of Earth.
The above-mentioned investigations have been instrumental in advancing the understanding of RE dynamics in the presence of wave–particle interactions in fusion plasmas. However, nearly all of these studies have been limited to slab geometry and have relied on quasilinear diffusion models. Crucially, these studies do not account for realistic 3D tokamak geometries, where discrete whistler eigenmodes emerge from coupling among multiple poloidal and radial harmonics. Incorporating these effects is essential for developing a comprehensive understanding of RE dynamics and for designing wave-based strategies to control runaway electrons in fusion plasmas.
% Several observational and numerical studies have reported that charged particles deviate from  a normal diffusion behavior due to wave-particle interactions in heliospheric and space plasma environements \onlinecite{Perri2008, Trotta2011, Allanson2019}. 

In this work, we investigate the physics of the RE-whistler interactions in tokamak plasmas using a first-principles-based modeling approach that couples the full wave-solver AORSA \cite{aorsa} (All-Orders Spectral Algorithm) with a tracer-particle code KORC \cite{carbajal,korc} (Kinetic Orbit Runaway Electrons Code). This approach enables the study of RE dynamics in realistic tokamak geometries by resolving discrete whistler eigenmodes and following full-orbit runaway electron trajectories in prescribed wave fields. We apply statistical analyses to characterize RE dynamics in pitch-angle space. In particular, to quantify pitch-angle scattering, we analyze the temporal scaling of the statistical variance, $\sigma^2$, of pitch-angle displacements. 
Transport is classified as diffusive when the variance scales linearly with time, i.e., ${\sigma^2} \sim t$, and as non-diffusive when the scaling follows $ {\sigma^2} \sim t^{\alpha}$ , with $\alpha \neq 1$. The latter corresponds to super-diffusive transport for $\alpha>1$ and sub-diffusive transport for $\alpha<1$, see e.g., Ref.~[\onlinecite{Diego2005}] and [\onlinecite{del_Castillo_Negrete_2006}].
Within this framework, we identify energy-dependent deviations from diffusive RE pitch-angle transport and provide numerical evidence of non-diffusive pitch-angle dynamics in the presence of whistler fields. These results highlight the role of wave-particle interactions in shaping RE transport and extend the description of RE transport in fusion plasmas beyond quasilinear diffusion models, such as the fractional diffusion model reported in Ref. [\onlinecite{Casolari_2024}] for representing sub-diffusive radial transport of REs in presence of small amplitude MHD perturbations in COMPASS tokamak.  
 
Several authors have discussed the phenomenon of anomalous diffusion of particles in complex physical systems \cite{ZASLAVSKY2002461,METZLER20001,Sanchez}. Previous studies~\cite{Allanson2019} reported anomalous pitch-angle diffusion in space plasmas using particle-in-cell simulations with simplified geometries. Resonant electron interactions with incoherent, high-amplitude waves were studied \cite{Frantsuzov_2023} and it was reported that the resulting diffusivity scales linearly with the wave-field amplitude, in contrast to the quadratic scaling predicted by quasilinear theory. An alternate formulation of the quasilinear diffusion models for wave-particle interactions in a relativistic plasma with axisymmetric non-uniform magnetic fields was given in Ref. [\onlinecite{Brizard_2022}]. However, these studies have limited applicability to fusion plasmas, where whistler waves involve coupling among multiple poloidal and radial harmonics in toroidal geometry, resulting in complex interactions with relativistic runaway electrons.
%Our framework captures anomalous diffusion of runaway electrons in both pitch angle and kinetic energy space, driven by resonant interactions with discrete whistler eigenmodes, and enhances the understanding of RE-whistler wave interactions in fusion plasmas beyond quasilinear approximations. 
%  
% AORSA  is a spectral full-wave solver that computes plasma eigenmodes without imposing any constraints on the wavelength relative to orbit size or on the number of cyclotron harmonics. KORC simulates the full-orbit runaway electron dynamics in the presence of external electromagnetic fields that are either described analytically or numerically calculated using other simulation codes based on tokamak geometry. We use AORSA to calculate whistler eigenmodes in a tokamak equilibrium and feed these fields into KORC, which then evolves a population of REs initialized with a spatially uniform distribution.
 
The rest of the manuscript is organized as follows. Section 2 describes the simulation model with a description of the dynamical equations solved in both AORSA and KORC codes. Section 3 describes the initial conditions and simulation setup used in the AORSA and KORC runs. Sections 4–6 present the main results obtained with the coupled AORSA–KORC framework, together with their physical interpretation. Finally, Section 7 summarizes the main findings of this study. 

\section{Computational model}
The individual codes used in the AORSA-KORC framework are described below and in greater detail in the cited references:
\subsection{AORSA}
The All-Orders Spectral Algorithm (AORSA) is a frequency-domain full-wave solver for calculating electromagnetic wave propagation and absorption in multidimensional toroidal plasmas~\cite{aorsa}. In the present work, AORSA is used to calculate the electromagnetic fields associated with whistler waves in the tokamak plasma. For a prescribed wave frequency $\omega$, AORSA solves the frequency-domain wave equation,
\begin{equation}
\nabla\times\nabla\times\mathbf{E}
-\frac{\omega^2}{c^2}\mathbf{E}-i\omega\mu_0\left(\mathbf{J}_{p}+\mathbf{J}_{\rm ant}\right)=0
\end{equation}
where $\mathbf{E}$ is the electric field, $\mathbf{J}_{\rm ant}$ is the externally specified antenna current, and $\mathbf{J}_{p}$ represents the plasma response to the wave fields. The plasma current is related to the wave electric field through the hot-plasma conductivity tensor, which incorporates finite-Larmor-radius effects without assuming $k_\perp\rho\ll1$; therefore, the plasma response is retained to all orders in $k_\perp\rho$. The formulation assumes that the equilibrium plasma varies slowly over a particle Larmor radius, i.e., $\rho/L\ll1$, where $L$ is the characteristic equilibrium scale length. Given the magnetic equilibrium, plasma density and temperature profiles, species composition, wave frequency, and antenna/source specification, AORSA calculates the self-consistent electric and magnetic fields throughout the plasma. In the present work, the three-dimensional electromagnetic fields calculated by AORSA are used by KORC to evolve the trajectories of runaway electrons and determine their interaction with the whistler wave.

\subsection{KORC}
The Kinetic Orbit Runaway electron Code (KORC) is a particle-based code developed to model the dynamics of relativistic runaway electrons in prescribed electromagnetic fields~\cite{carbajal,korc}. KORC can evolve both full-orbit (FO) and guiding-center (GC) trajectories of REs in general three-dimensional magnetic and electric fields. In the full-orbit formulation used in this work, the position and momentum of each RE are evolved according to the relativistic equations of motion,
\begin{equation}
\frac{d\mathbf{x}}{dt}= \mathbf{v}; \ \mathbf{p}= {\gamma m \mathbf{v}}, \qquad
\frac{d\mathbf{p}}{dt}
=-e\left(\mathbf{E}+\mathbf{v}\times\mathbf{B}\right),
\end{equation}
where $\mathbf{x}$, $\mathbf{v} $ and $\mathbf{p}$ are the particle position, velocity and relativistic momentum, respectively, $\gamma$ is the relativistic Lorentz factor and $m$ is the rest mass of the tracer particle. Given an initial RE distribution and prescribed equilibrium and perturbed electromagnetic fields, KORC follows individual RE trajectories and calculates their evolution in configuration and momentum space.
\par
In the present work, the electromagnetic fields of the whistler waves calculated using AORSA are provided to KORC in addition to the equilibrium tokamak magnetic fields (generated using EFIT \cite{Fitzgerald_2013}). KORC then evolves an ensemble of 10,240 full-orbit RE trajectories in the combined equilibrium and wave fields. This particular number of 10,240 REs was chosen to maximize particle statistics by utilizing the available parallel computation resources without increasing the wall-clock simulation time. The resulting particle trajectories are used to quantify changes in RE kinetic energy and pitch angle and to characterize the transport produced by RE--whistler wave interactions.

%KORC has previously been used to analyze runaway electron physics in several tokamak devices, such as DIII-D \onlinecite{Hollmann_2025}, MST \onlinecite{cornille2022} and JET \onlinecite{Beidler21}.
%  
% Figure \ref{fig:1} shows the EFIT equilibrium from DIII-D shot $\#171089$, during which runaway electron-driven whistler waves were observed \onlinecite{spong,heidbrink2019}.
\section {Simulation setup}
\begin{figure*}[htb!]
    \centering
    \includegraphics[height=12cm,width=13cm]{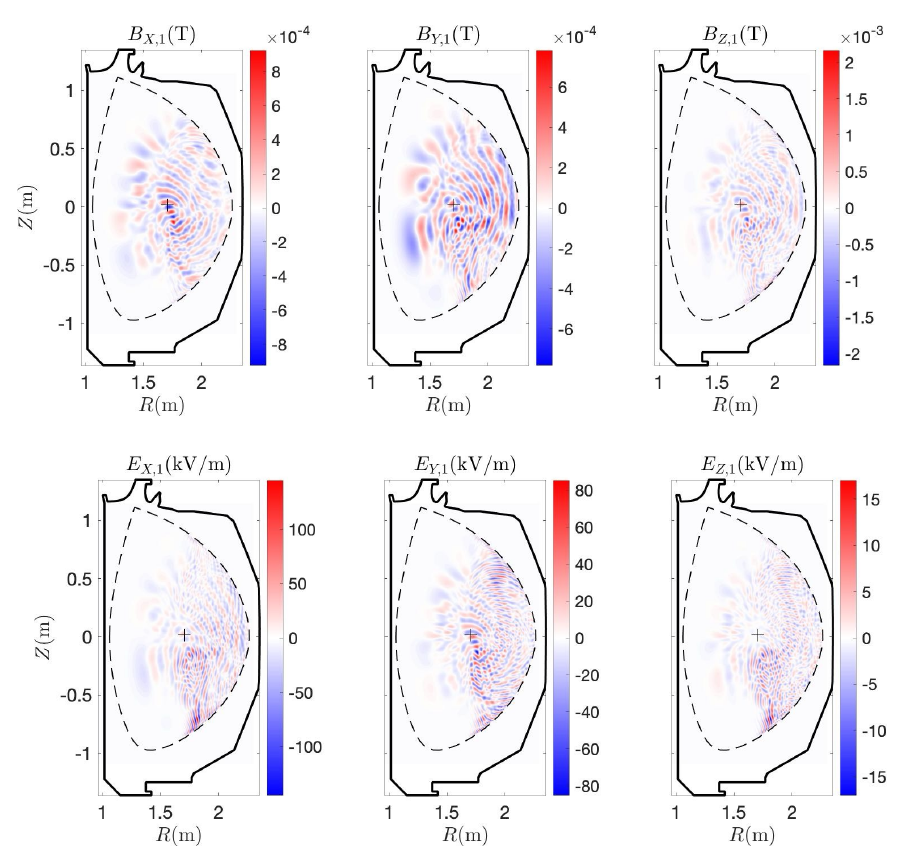}
    \caption{Magnetic field (top) and electric field (bottom) fluctuations corresponding to whistler waves calculated by AORSA for a DIII-D type plasma equilibrium. }
    \label{fig:3}
\end{figure*}

For evolving RE trajectories, the tokamak plasma equilibrium provided to KORC corresponds to the DIII-D whistler-wave experiments~\cite{spong,heidbrink2019}, together with the associated density and temperature profiles shown in Fig.~3 of Ref.~[\onlinecite{heidbrink2019}]. Using these inputs, AORSA computes whistler fields at a prescribed antenna frequency of 200 MHz and toroidal mode number n=35, consistent with the wave-parameters observed in the DIII-D whistler–RE experiments.  For the plasma parameters considered, the prescribed 200 MHz frequency lies on the whistler branch of the cold-plasma dispersion relation.
The resulting wave structures, illustrated in Fig.~\ref{fig:3}, exhibit strong coupling among multiple poloidal and radial harmonics, as well as mode localization on the low-field side due to fast-wave cutoff conditions being satisfied in the high-field-side region.
%\textbf{In the absence of experimental measurements of whistler-wave amplitudes in DIII-D tokamak, we adopt representative (but not experimentally constrained) perturbation amplitudes for whistler fields. We limit the maximum normalized magnetic field amplitude to $\delta B_{max}/B_0\approx10^{-3}$, which is consistent with the order-of-magnitude upper bound of normalized electromagnetic fluctuations typically observed in tokamak plasmas.} 

% \begin{figure}[htb]
%     \centering
%     \includegraphics[height=6cm,width=8cm]{EFITRZ.jpg}
%     \caption{Equilibrium flux surface plots for DIII-D shot $\#171089$. }
%     \label{fig:1}
% \end{figure}

%\begin{figure}[h]
%    \centering
 %   \includegraphics[height=6cm,width=13cm]{scaled_B_fields.jpg}
  %  \caption{Magnetic field  fluctuations corresponding to whistlers waves calculated by AORSA for a DIII-D type plasma equilibrium. }
   % \label{fig:3}
%\end{figure}
%\begin{figure}[h]
 %  \centering
%    \includegraphics[height=6cm,width=13cm]{scaled_E_fields.jpg}
 %   \caption{Electric field fluctuations corresponding to whistlers waves calculated by AORSA for a DIII-D type plasma equilibrium.}
  %  \label{fig:2}
%\end{figure}

\par The AORSA-KORC framework is a one-way coupling where runaway electrons evolve in prescribed wave fields and the RE dynamics do not impact whistler fields in any way, and hence, the whistler time period ($T_w$) is constant for prescribed wave-fields calculated at a particular frequency of 200 MHz using AORSA.
\par For a single plane wave, resonant particles may be identified using the relativistic resonance condition
    \begin{equation}
    \omega - k_{\parallel}v_{\parallel}
    - \frac{\ell\Omega_{ce}}{\gamma} = 0. 
    \label{eq:res}
    \end{equation} 
    In the present calculation, however, the whistler eigenmode calculated with AORSA in a 3D tokamak geometry (Fig.~\ref{fig:3}) is a global toroidal eigenmode. For this global eigenmode solution, the parallel wave number can be approximated as $k_{\parallel} \simeq(n-m/q)/R$, where, 'n' and  'm' are, respectively the toroidal and poloidal mode numbers, 'q' is the safety factor which is a measure of the helicity of the magnetic field and 'R' is the major radius location. Although the toroidal mode number $n=35$ and frequency $f=200$~MHz are prescribed, the eigenmode contains multiple coupled poloidal harmonics whose relative coupling amplitudes vary with radial position, due to variation of both 'q' and 'R', and the changing poloidal modes result in different local values of $k_{\parallel}$ throughout the plasma. Consequently, there are multiple $k_{\parallel}$ that characterize the wave throughout the plasma.

 \par Due to mode couplings, wave-particle interactions in the tokamak geometry are considerably more complex than a slab geometry and differ significantly from the simplified single-mode resonance condition such as Eq. (\ref{eq:res}). Rather, multiple resonant wave numbers may interact simultaneously with runaway electrons of different energies present in the fusion plasma, producing a richer and more complex phase-space response, which is captured by our AORSA-KORC simulation model. In order to analyze how these complex interactions impact the runaway electron transport in the pitch angle-kinetic energy space, we use statistical analysis as a tool to uncover the underlying physics.

% To reduce computational costs while maintaining physical fidelity, the AORSA-derived whistler fields (both electric and magnetic fields) were scaled by a factor of 100, resulting in peak values of $\delta B_{max}=2.19*10^{-3}$\ T, $\delta E_{max}=160 \ \text{kV/m}$. This scaling is used because power transfer from waves to particles is proportional to the product of wave electric field and the wave-particle interaction time. Thus, simulating an increased wave electric field over a shorter duration approximated the effect of a longer simulation with unscaled fields. Throughout this manuscript, "whistler fields" refer to these scaled AORSA fields.  
%% Simulation set up 
% \section{Simulation setup}
%We begin our simulations by running KORC to obtain RE distributions close to the physically realistic distributions in fusion experiments. We then simulate RE dynamics in the presence of whistler fields calculated using the AORSA code for the DIII-D model equilibrium where whistler waves were observed \onlinecite{spong}. Finally, we conducted statistical analysis on the simulation results. 
  
To construct a physically consistent initial condition, we integrated the trajectories of an ensemble of 10,240 REs for $1 \ \text{ms}$ in the absence of whistler-wave fields and collisions, with energies uniformly distributed in the interval $1-20 \ \text{MeV}$ and an initial pitch angle of $\eta=10^\circ$, where the pitch angle is defined as 
$\eta = \tan^{-1}\left(v_\perp/v_{\parallel}\right)
$,
where $v_{\parallel}$ and $v_\perp$ are the velocity components parallel and perpendicular to the local magnetic field. The REs were initialized uniformly in the $R$--$Z$ plane within the plasma and in the toroidal direction to provide broad spatial sampling of the whistler-wave field. While the local interaction strength depends on radial position due to the spatial variation of the wave amplitude, the results presented here are ensemble statistics over the full spatially distributed RE population rather than a particular radial population.
\par Since the pitch angle is not a constant of motion, the initial mono-pitch distribution undergoes collisionless pitch-angle scattering~\cite{carbajal} and relaxes toward a quasi-steady pitch angle distribution characterized by a broader range of pitch-angles spanning from $0^\circ- 60^\circ$. This relaxed, physically consistent distribution was then used as the initial condition to investigate the effect of the AORSA-computed whistler-wave fields on the RE ensemble over further $2 \ \text{ms}$ time scale.
\par
The time step for KORC full-orbit simulations is chosen to be
$\Delta t = 0.01T_{ce,\mathrm{rel}}$, corresponding to approximately $100$ time steps per relativistic electron gyroperiod $T_{ce,\mathrm{rel}} \simeq 5.94\times10^{-11}$~s calculated for a $1$~MeV electron at $B=1.77$~T. The prescribed $200$~MHz whistler wave has a period of $T_w=5$~ns; therefore, the $2$~ms interval ($1<t<3$~ms) over which the whistler fields are included corresponds to approximately $4\times10^{5}$ whistler-wave periods of RE--whistler interactions.
\section{Non-Gaussian statistics of pitch angle scattering }
\begin{figure*}[htb!]
    \centering
    \begin{subfigure}{0.45\textwidth}
        \includegraphics[width=\linewidth]{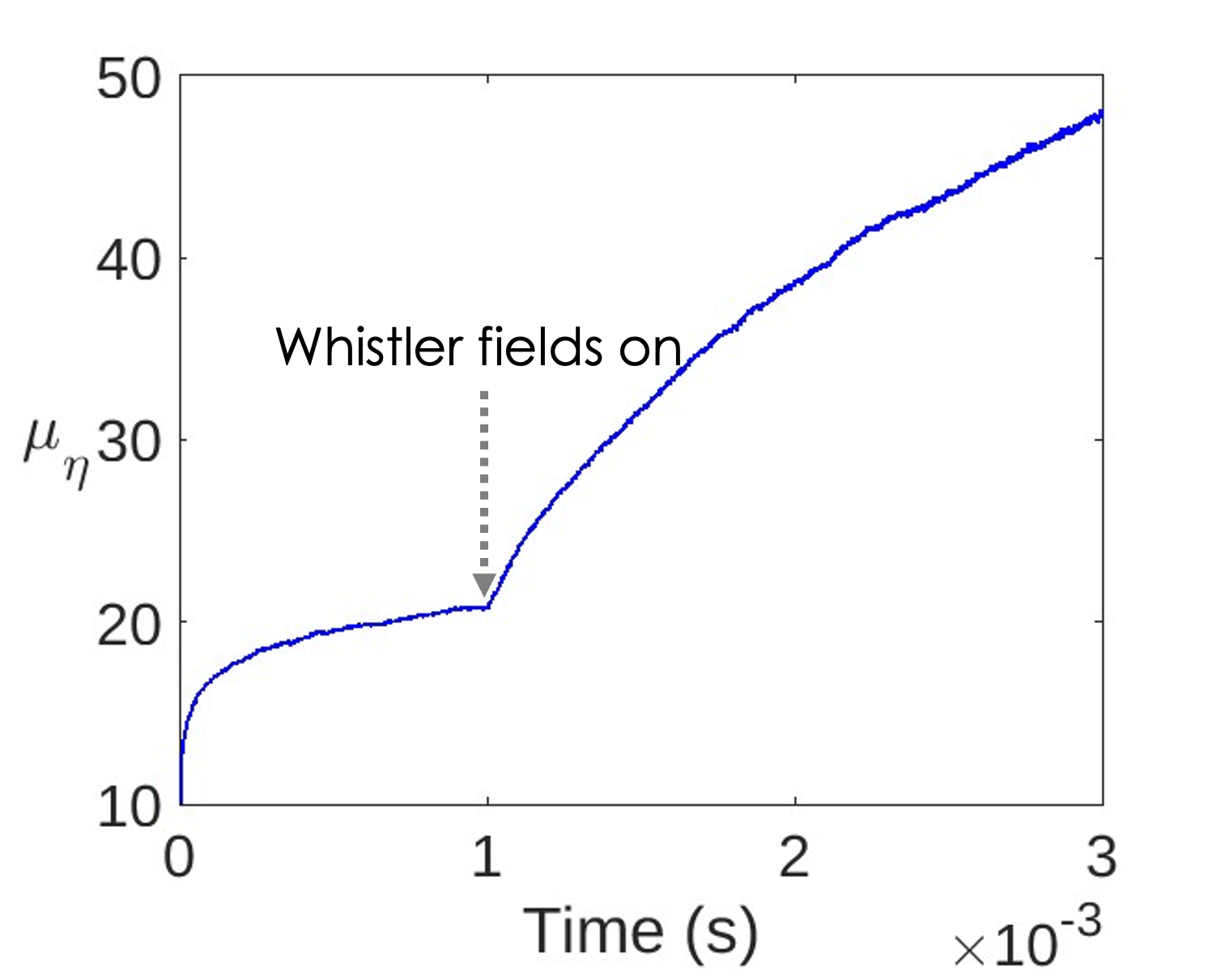}
        \caption{}
        \label{subfig1eta}
    \end{subfigure}
    \hfill
    \begin{subfigure}{0.45\textwidth}
        \includegraphics[width=\linewidth]{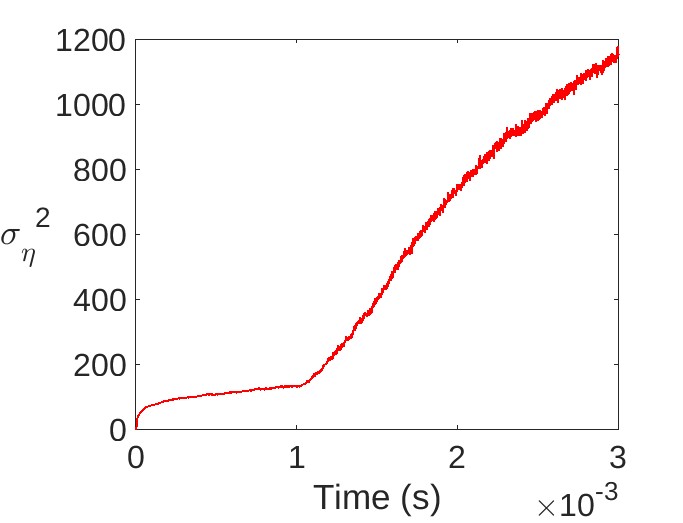}
        \caption{}
        \label{subfig2eta}
    \end{subfigure}
    \vspace{0.5cm}
    \begin{subfigure}{0.45\textwidth}
        \includegraphics[width=\linewidth]{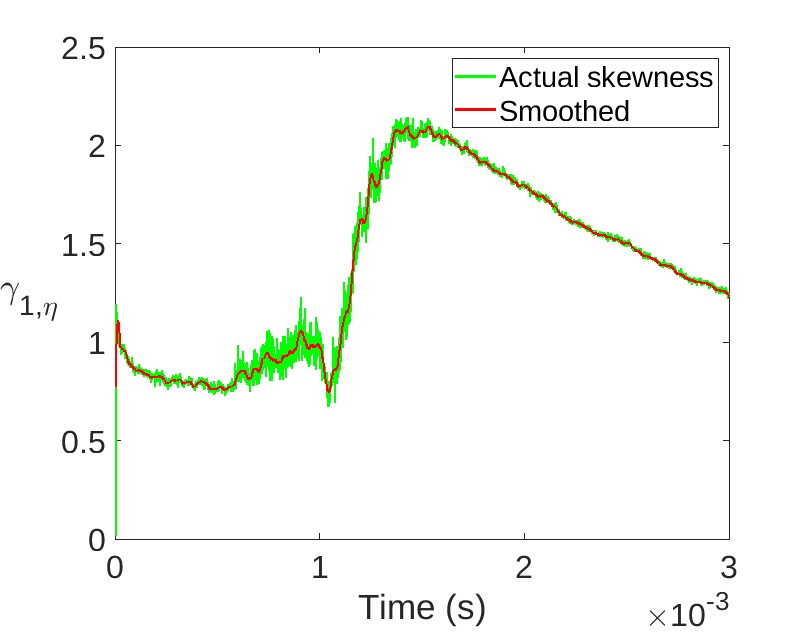}
        \caption{}
        \label{subfig3eta}
    \end{subfigure}
    \hfill
    \begin{subfigure}{0.45\textwidth}
        \includegraphics[width=\linewidth]{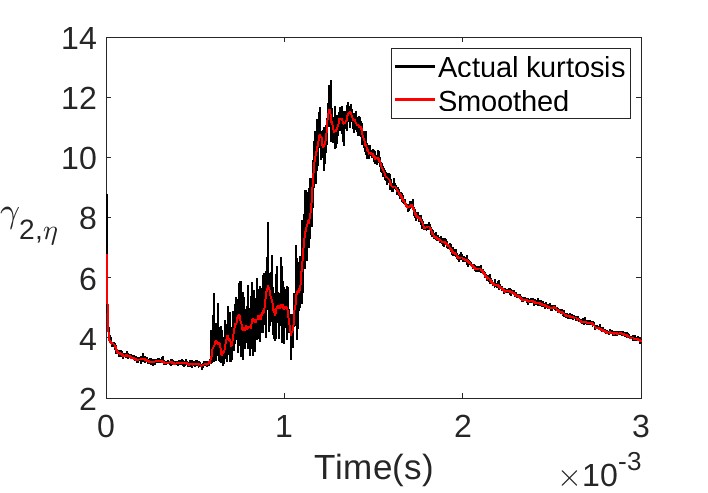}
        \caption{}
        \label{subfig4eta}
    \end{subfigure}
    \caption{Time evolution of the statistical moments of the runaway electron pitch angle distribution. The time interval $[0,1]$ms corresponds to the collisionless pitch-angle-scattering relaxation of the distribution in the absence of whistler waves. During the interval $[1,3]$ms, the REs evolve in the presence of a whistler field with $\delta B_{\text{max}}/(B_0) = 10^{-3}$. In this interval, the statistical moments of the distribution exhibit a rapid increase, followed by a relaxation phase in the case of the third and fourth moments.}
    \label{moments_eta}
\end{figure*}

To demonstrate the effect of whistler fields on the runaway electron ensemble, in Fig. \ref{moments_eta} we show the time evolution of the first four statistical moments of the pitch angle $\eta$ distribution, defined as: 
\begin{equation}
 \mu_\eta = \langle \eta \rangle \, \qquad
 \sigma_\eta^2 = \langle (\eta - \mu_\eta)^2 \rangle\, , 
 \end{equation}
 \begin{equation}
 \gamma_{1,\eta} = \frac{\langle (\eta - \mu_\eta)^3 \rangle}{\sigma_\eta^3}\, \qquad
 \gamma_{2,\eta} = \frac{\langle (\eta - \mu_\eta)^4 -3 \rangle}{\sigma_\eta^4}
\end{equation}
%\begin{align} 
%\mu_\eta &\equiv \langle \eta \rangle, \nonumber \\ 
%\sigma_\eta^2 &\equiv \langle (\eta - \mu_\eta)^2 \rangle, \nonumber\\ 
%\gamma_{1,\eta} &\equiv \frac{\langle (\eta - \mu_\eta)^3 \rangle}{\sigma_\eta^3}, \nonumber \\ 
%\gamma_{2,\eta} &\equiv \frac{\langle (\eta - \mu_\eta)^4 -3 \rangle}{\sigma_\eta^4}, 
%\end{align}
where $\langle \cdot \rangle$ denotes ensemble average at a given time. 
Figures \ref{moments_eta} (a) and (b) display the mean, $\mu_\eta$, and variance, $\sigma^2_\eta$ respectively, while Figs.~ \ref{moments_eta} (c) and (d) show the normalized third, $\gamma_{1,\eta}$, and fourth, $\gamma_{2,\eta}$, central moments (skewness and kurtosis). 
  During the initial collisionless pitch angle scattering, 
$t \leq 1\ \text{ms}$, $\mu_\eta$ increases rapidly over $0.1 \ \text{ms}$ and then subsequently saturates, varying by less than $ 10\%$ until $t=1 \ \text{ms}$. Over the same interval, $\sigma^2_\eta$ remains relatively small, indicating limited pitch angle dispersion in the absence of electromagnetic fluctuations. At $t=1 \ \text{ms}$, whistler fields computed by AORSA are introduced into KORC simulations. Following the onset of whistler fields ($1 \leq t \leq 3 \ \text{ms}$), both the mean and variance increase monotonically, demonstrating enhanced pitch-angle scattering of REs due to interactions with whistler eigenmodes. Such scattering of REs to high pitch angles implies an increase in the perpendicular velocity component, which, in principle, can enhance synchrotron radiation losses. While a quantitative evaluation of synchrotron radiation damping resulting from this pitch-angle scattering is beyond the scope of the present work, these results suggest a potential pathway by which whistler-induced scattering may contribute to RE mitigation in future fusion devices.
 
The evolution of higher-order moments further highlights this transition to enhanced pitch angle scattering regime with onset of whistlers. As shown in Fig.~\ref{moments_eta} (c), the skewness attains a finite positive value shortly after $t=0$, reflecting the development of asymmetry in the initially mono-pitch distribution centered at $10^{\circ}$, with a preferential spread towards larger pitch angles. After the onset of whistler fields, the skewness increases further and then decreases after $1.5 \ \text{ms}$, indicating a gradual relaxation towards a more symmetric distribution, but with a larger mean pitch angle.
 
\begin{figure*}[htb!]
    \centering
    \includegraphics[width=1.0\linewidth]{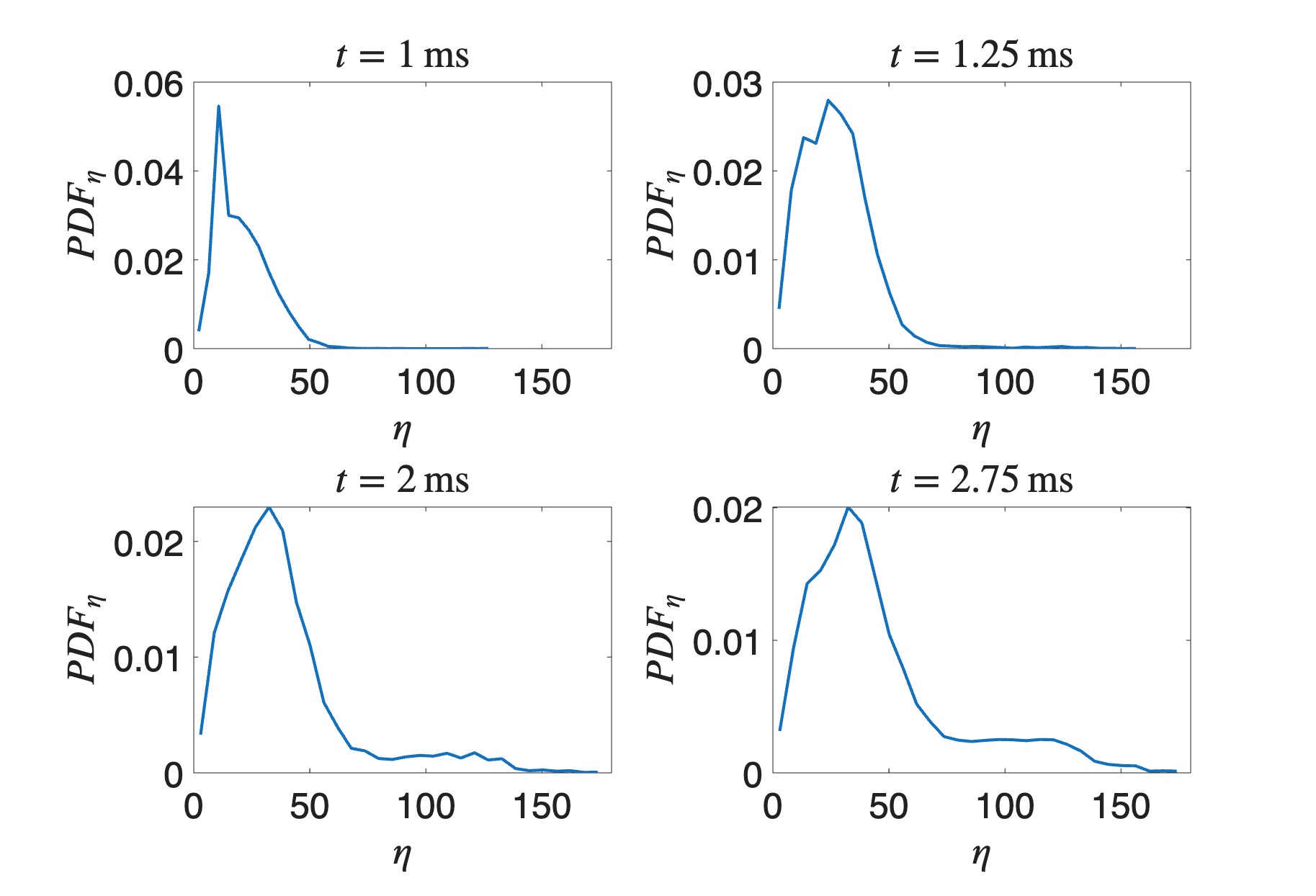}
    \caption{Probability distribution functions (PDFs) of the RE pitch angle distributions at different simulation times.}
    \label{PDF_eta}
\end{figure*}

Figure~\ref{moments_eta}(d) shows the time evolution of the kurtosis, which measures the propensity of a distribution to produce rare, large deviations from the mean. Following the introduction of the whistler fields, the kurtosis increases rapidly, indicating that the pitch-angle distribution develops heavy tails and exhibits extreme outliers relative to a Gaussian distribution. These extreme events arise from rare but strong pitch-angle scattering events of REs induced by the whistler fields.
The subsequent decrease in the kurtosis indicates a transition toward less intermittent dynamics, characterized by a reduced occurrence of rare extreme events and the absence of heavy tails as the high pitch angle regions of the distribution get increasingly populated. This is further explained in Fig. \ref{PDF_eta}.

Going beyond the analysis of the statistical moments, Fig.~\ref{PDF_eta} shows the PDFs (probability distribution functions) of the RE pitch-angle distribution at selected times during the simulation. 
The PDF at $1\ \text{ms}$, corresponds to the relaxed, quasi-steady pitch angle distribution in the absence of the whistler field. The sharp peak at $\eta = 10^\circ$ reflects the concentration of REs near their initial pitch angle before the whistler fields are introduced. Following the introduction of whistler fields, the pitch angle distribution broadens progressively, showing an increasing RE population at larger pitch angles. In particular, from $t=2\ \text{ms}$ onward, the development of an extended high-pitch-angle tail is observed, consistent with enhanced pitch angle scattering due to wave-particle interactions. This evolution is consistent with the corresponding increase in the mean and variance, as well as the transient enhancement in skewness and kurtosis, discussed in Fig.~\ref{moments_eta}. 
\par
At later times, the REs increasingly populate the high pitch angle regions and the mean and width of the distribution increase (Figs.~\ref{moments_eta}(a) and \ref{PDF_eta}(c)--(d)). Consequently, the tail becomes less exceptional relative to the broadened distribution, resulting in the decrease in kurtosis and skewness between 2-3 ms even though the population at large pitch angles continues to increase.

%PDFs for eta  
 
\begin{figure}
    \centering
    \includegraphics[width=1.0\linewidth]{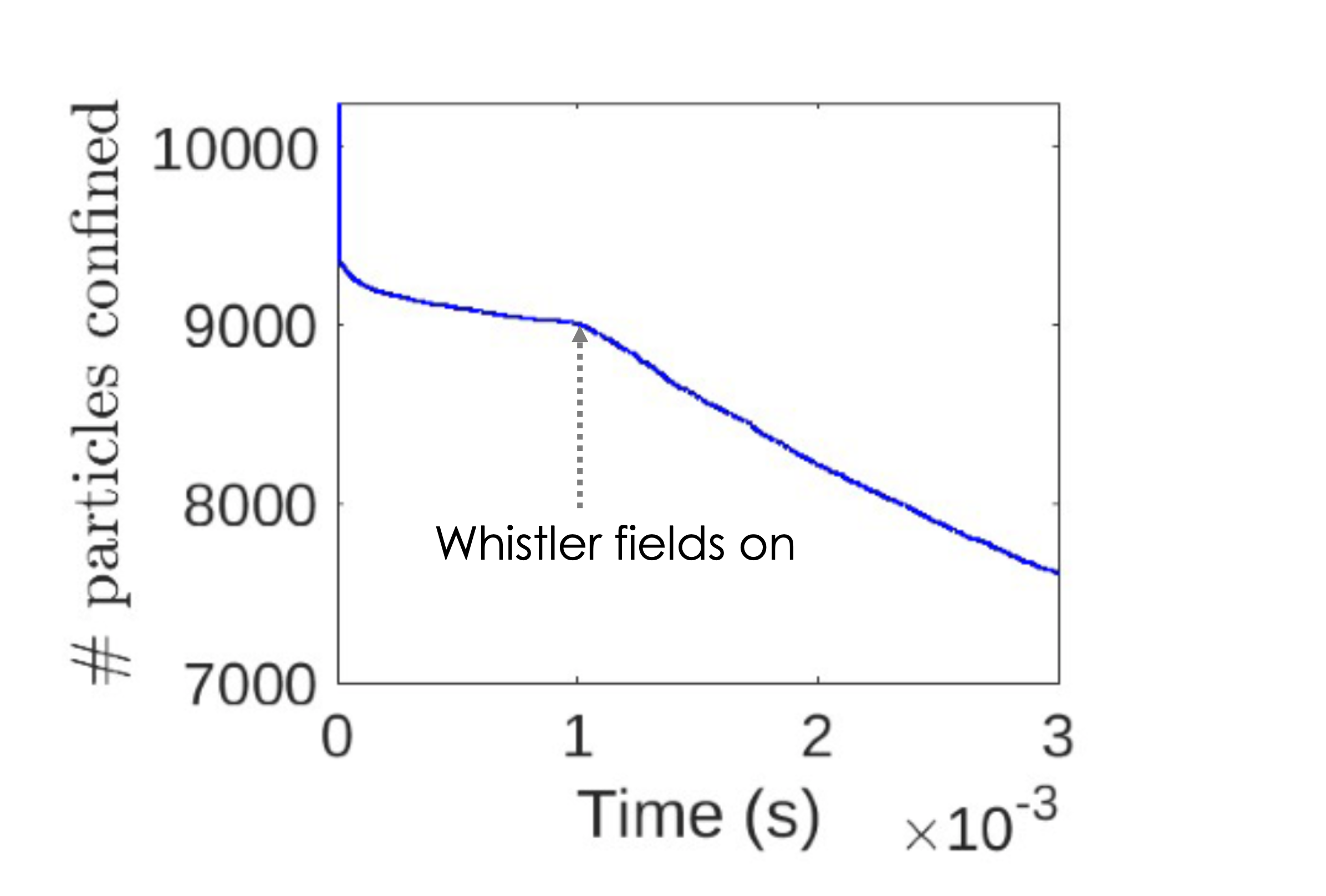}
    \caption{Total number of confined runaway electrons (REs) as a function of time during the simulation. }
    \label{confined_particles}
\end{figure}

It is important to note that in these simulations, the number of confined particles does not stay constant. As Fig.~\ref{confined_particles} shows, there is a prompt loss of about $700$ particles at the beginning of the simulation. 
Following the prompt loss phase, the number of confined particles reaches a plateau and varies by less than $4\%$ during the interval $[0,1]$ms. However, once the whistler fields are switched on, $t>1$ms, the number of confined particles exhibits a rapid linear decay. This trend may be attributed to an increase in RE pitch angle of the REs that leads to larger RE gyro-radii and displacements from initial flux surfaces, causing greater RE losses, especially near the last closed flux surface.
  
While the PDFs in Fig.~\ref{PDF_eta} show the cumulative effect of pitch-angle scattering, they do not directly reveal the statistical nature of individual scattering events. To address this, we analyze the PDF of pitch angle increments, or ``kicks" defined as $\Delta \eta (t)=\eta(t+T_w)-\eta(t)$, where $T_w$ is the whistler time period ($T_w=5 \ \times 10^{-6}\ \text{ms}$). To obtain these distributions, KORC simulations are initialized using phase-space coordinates of REs sampled at $t=1.25 \ \text{ms}$, corresponding to $0.25 \ \text{ms}$ after the onset of whistler fields (see Fig.~\ref{moments_eta}). The REs are then evolved for a duration of $10^{-2} \ \text{ms}$ in the presence of whistler fields, with pitch angles recorded at intervals of one whistler period. 
  
\begin{figure*}[htb!]
    \centering
    \begin{subfigure}{0.45\textwidth}
        \includegraphics[width=\linewidth]{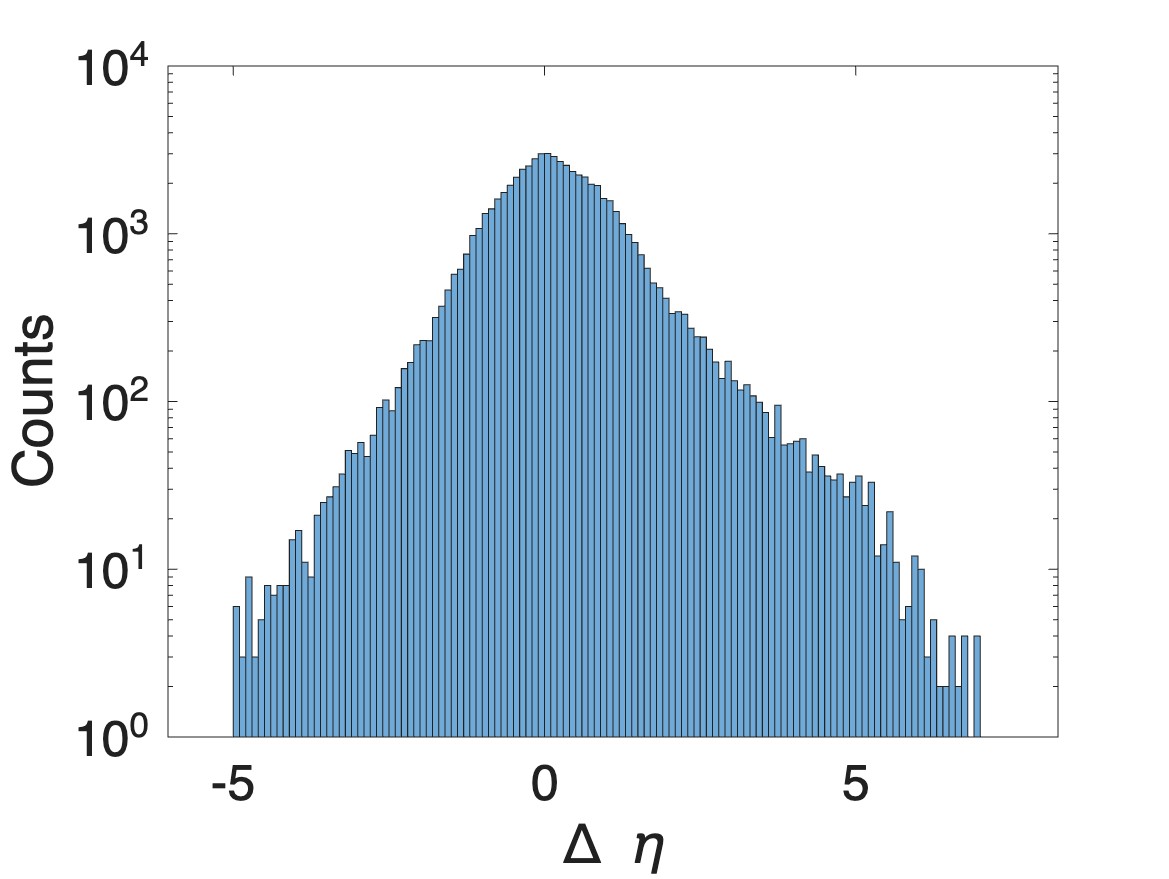}
        \caption{}
    \end{subfigure}
    \hfill
    \begin{subfigure}{0.5\textwidth}
        \includegraphics[width=\linewidth]{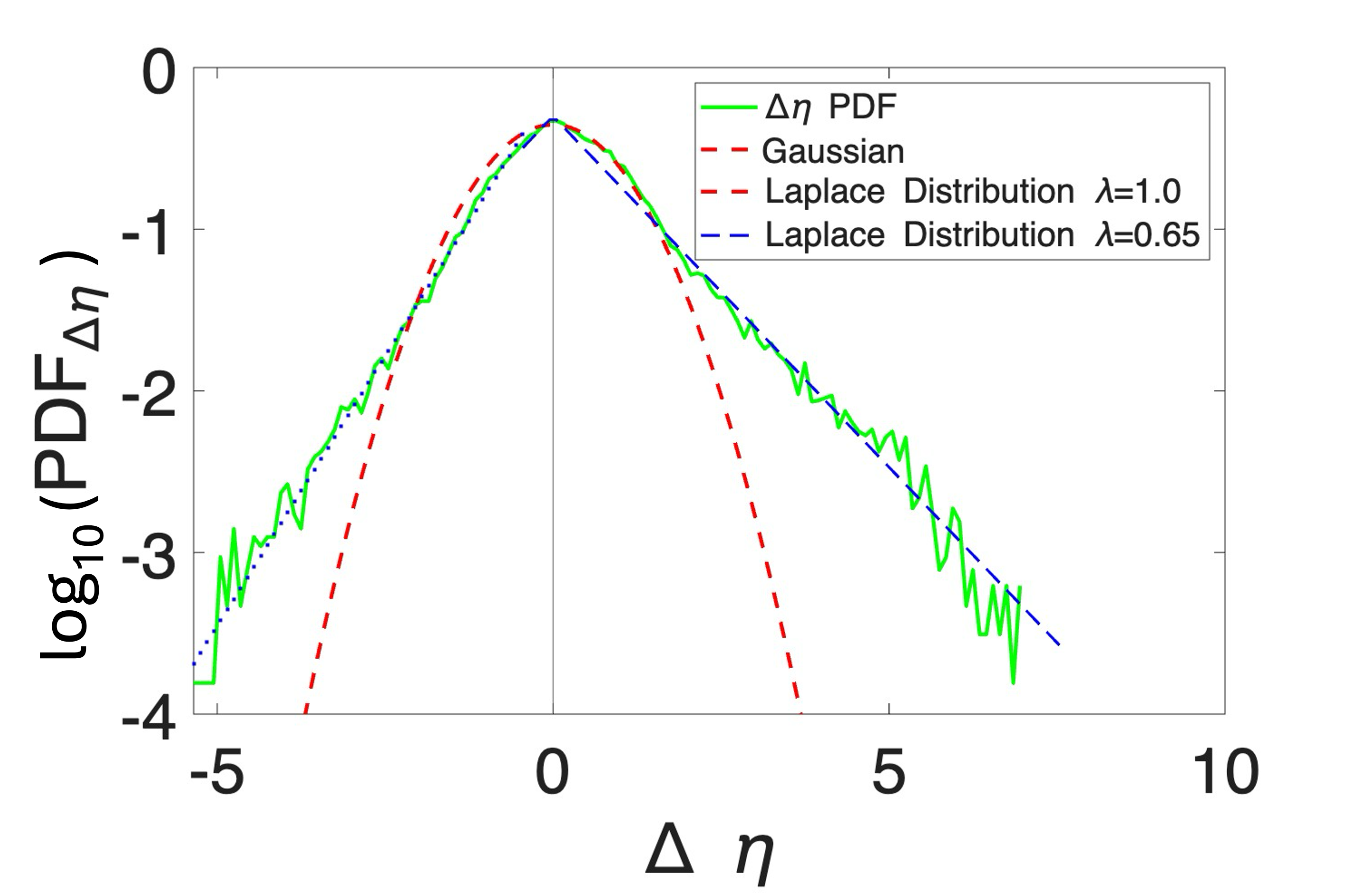}
        \caption{}
    \end{subfigure}
    \caption{Distribution of RE pitch angle kicks, $\Delta \eta$, obtained from KORC simulations initialized at time $t=1.25 \ \text{ms}$ and evolved for $10 \ \mu\text{s}$ in the presence of whistler fields and recorded at whistler time period of $5 \ \text{ns}$. a) Histogram of $\Delta \eta$ kicks for $\eta_{inst} \in [1^{\circ}-5^{\circ}]$ and $K_{inst} \in [8,10]\ \text{MeV}$ and b) Normalized probability distribution of $\Delta \eta$ kicks on log-log scale fitted by Gaussian and Laplace distributions for the same range of pitch angle and kinetic energies. }
    \label{PDF1}
\end{figure*}

Figure \ref{PDF1} (a) shows the resulting histogram of $\Delta \eta$ kicks obtained for REs with instantaneous pitch angles in the interval $[1^{\circ},5^{\circ}]$ and instantaneous kinetic energies in the interval $[8,10] \ \text{MeV}$. As shown in Fig.~\ref{PDF1} (b), the core of this $\Delta \eta$ distribution is well-approximated by a Gaussian, while the tails are  well-described by a Laplace distribution of the form: $\frac{1}{2 \lambda}\,\exp\!\left(-\frac{|\Delta\eta|}{\lambda}\right)$. The exponential decay of the distribution tails is controlled by the Laplace scale parameter, $\lambda$. Notably, different values of $\lambda$ are required to fit the positive and negative tails ($\lambda =1$ and $\lambda=0.65$, respectively), indicating an asymmetric distribution of pitch-angle increments. This asymmetry reflects a preferential increase in pitch angles for a subset of REs within this specific pitch angle and kinetic energy range, consistent with the presence of rare, large pitch angle excursions experienced by a subset of REs. Such intermittent scattering events provide a microscopic explanation for the enhanced skewness and kurtosis observed in Fig.~\ref{moments_eta}.
 
\begin{figure*}[htb!]
    \centering
    \begin{subfigure}{0.45\textwidth}
        \includegraphics[width=\linewidth]{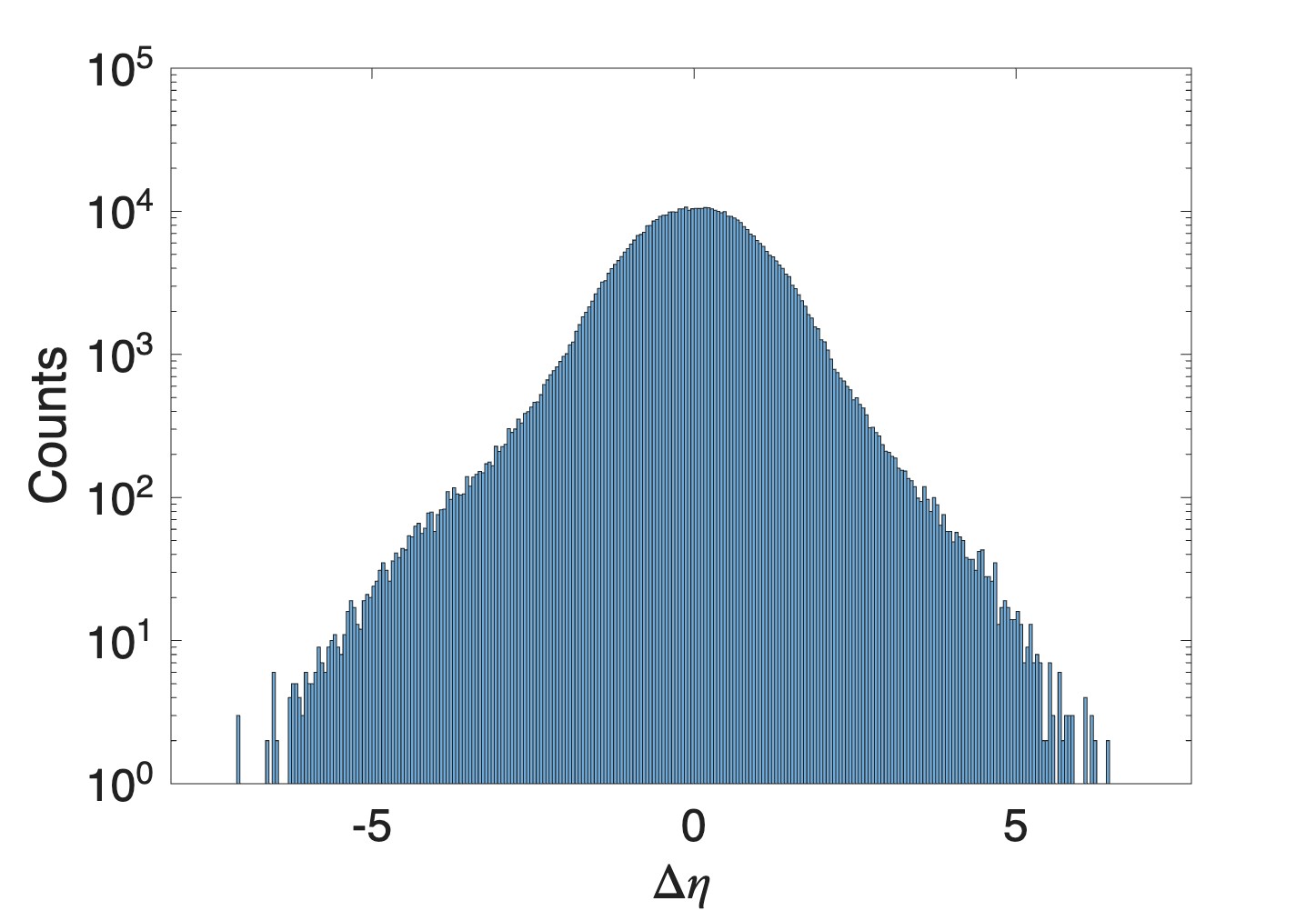}
        \caption{}
    \end{subfigure}
    \hfill
    \begin{subfigure}{0.5\textwidth}
        \includegraphics[width=\linewidth]{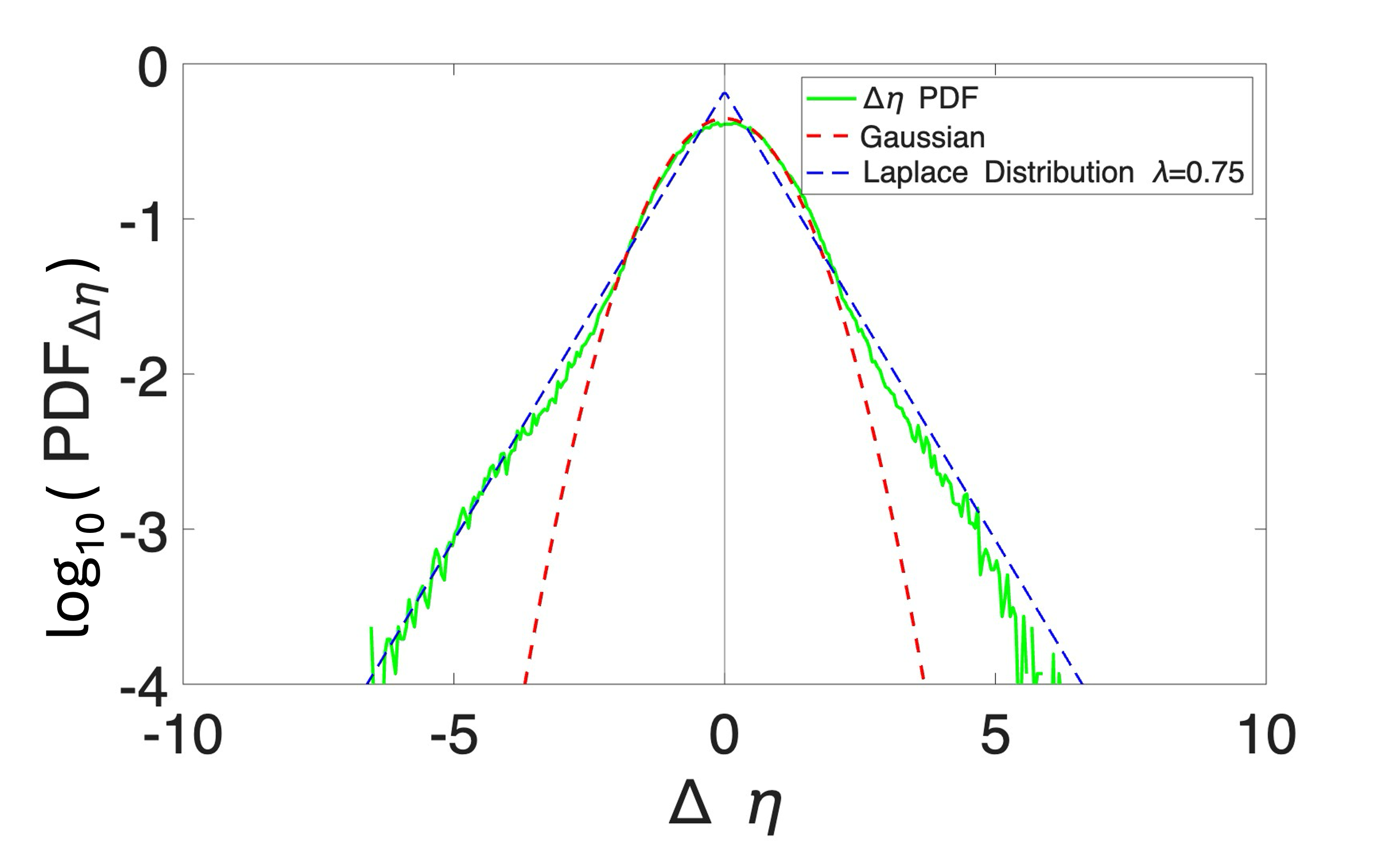}
        \caption{}
    \end{subfigure}
    \caption{Same as Figure \ref{PDF1} but for 
    $\eta_{inst} \in [6^{\circ},16^{\circ}]$ and $K_{inst} \in [8,9]\ \text{MeV}$.}
    \label{PDF2}
\end{figure*}

Figure \ref{PDF2} shows the $\Delta \eta$ distribution for a different phase-space region with instantaneous pitch angles 
$\eta_{inst} \in [6^{\circ},16^{\circ}]$ and $KE_{inst} \in [8,9]\ \text{MeV}$.
 While the variance of the distribution remains large, the $\Delta \eta$ distribution in this case is approximately symmetric, with both tails well-described by Laplace distributions with the same scale parameter $\lambda=0.75$.

 %The contrasts between figure \ref{PDF1} and highlights a strong dependence of pitch angle increment statistics on the initial pitch angle and kinetic energy of the RE population, consistent with the energy-dependent anomalous pitch-angle transport discussed in the following section. 
 The insights obtained from kick-statistics in Figs.~\ref{PDF1} and \ref{PDF2} might help develop transport probability distribution matrices for REs due to interactions with whistler waves, like the ones built for the kick-model of fast ions transport \cite{Mario_2014} in the presence of Alfvén eigenmode instabilities in fusion plasmas.

\section{Non-diffusive scaling and limitations of quasilinear diffusion model}
To characterize the long-time behavior of RE pitch-angle transport in the presence of whistler fields, we analyze the temporal scaling of the variance 
of pitch-angle displacements with respect to the initial condition,
${\sigma_{\delta \eta}}^2=<[\delta \eta-<\delta \eta>]^2>$, where 
$\delta \eta(t)=\eta(t)-\eta_0$.
%The variance of pitch angle displacements with respect to the initial condition is defined as 
%\begin{equation}
    %{\sigma_i}^2=<[\delta \eta_i-<\delta \eta_i>]^2>; \qquad \delta \eta_i =\eta_i-\eta_0; \qquad i \ \in \ [0,t]
%\end{equation}
The scaling of ${\sigma_{\delta \eta}}^2$ with time allows us to identify the presence of non-diffusive transport. When ${\sigma_{\delta \eta}}^2 \sim t^\alpha$ with $\alpha = 1$ transport is diffusive. However, when $\alpha >1$ transport is super-diffusive and when $\alpha <1$ transport is sub-diffusive, see  Ref.~\onlinecite{Diego2005} and references therein. 

\begin{figure*}[htb!]
    \centering
    \begin{subfigure}{0.45\textwidth}
        \includegraphics[width=\linewidth]{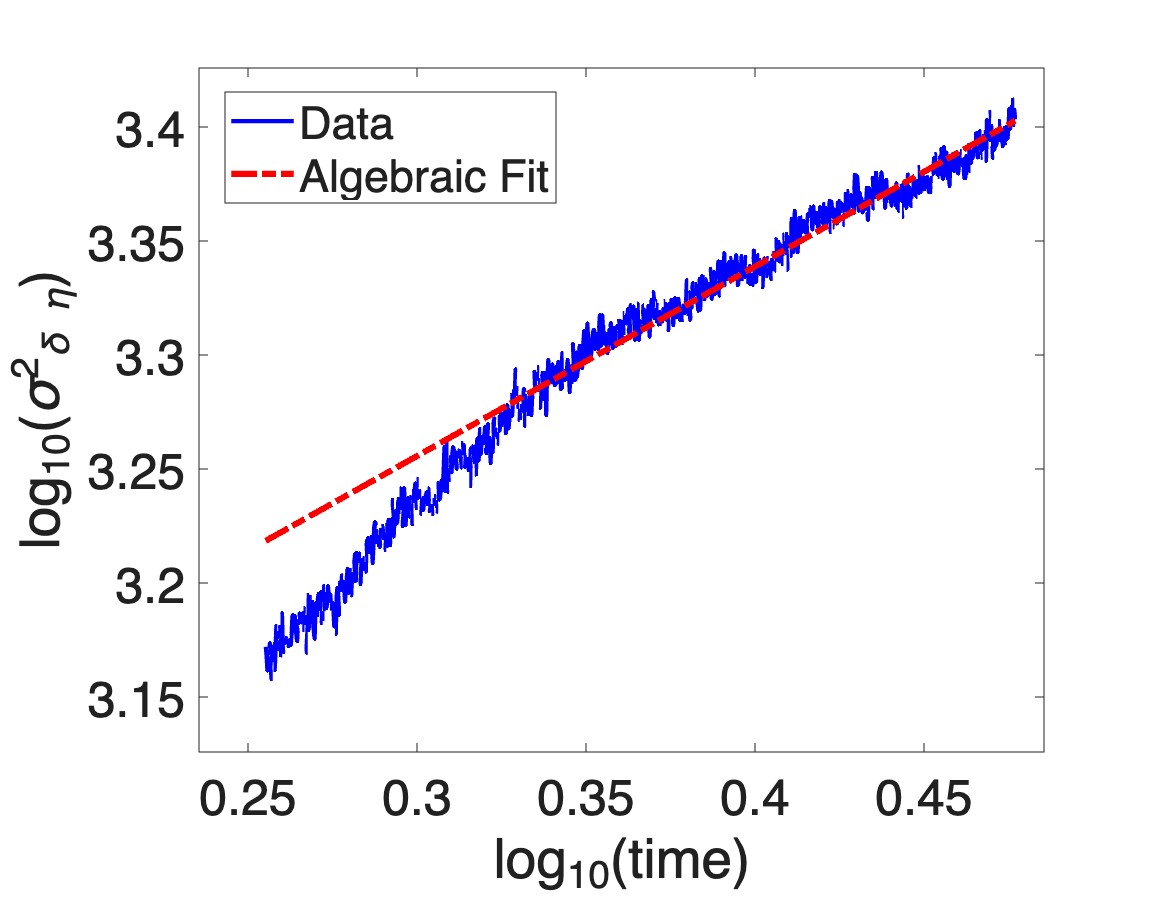}
        \caption{$K_{0}= 1-5 \ \text{MeV}$, $\sigma^2 \approx t^{0.75}$}
        \label{subfig1a}
    \end{subfigure}
    \hfill
    \begin{subfigure}{0.45\textwidth}
        \includegraphics[width=\linewidth]{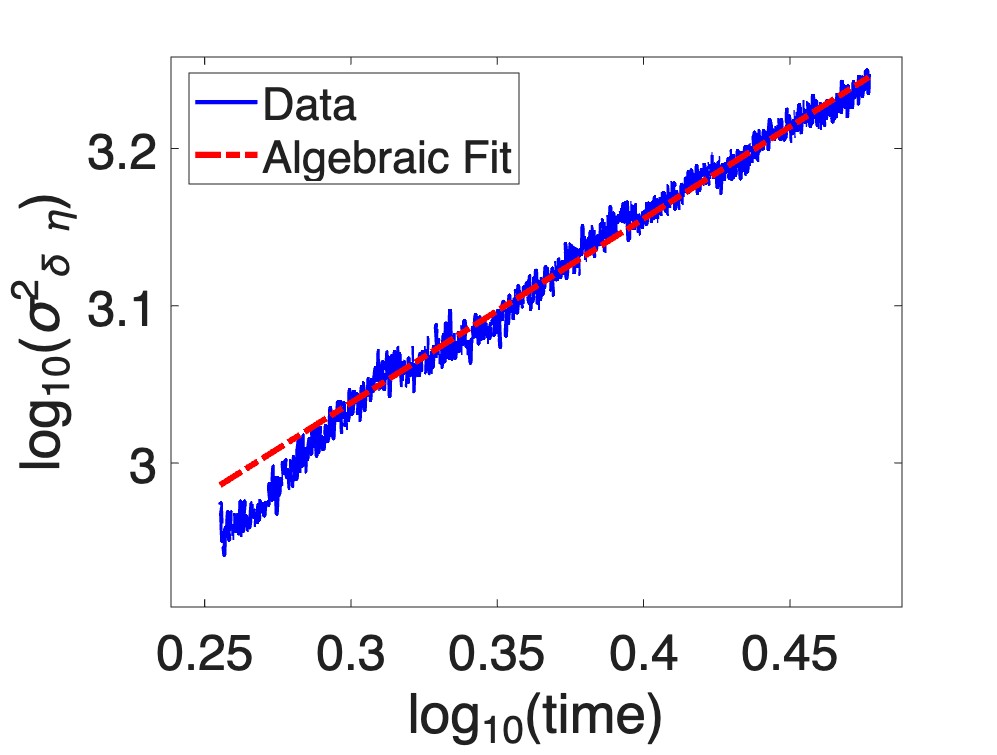}
        \caption{$K_{0}= 5-10 \ \text{MeV}$,$\sigma^2 \approx t^{1.2}$}
        \label{subfig1b}
    \end{subfigure}
    \\
    \begin{subfigure}{0.45\textwidth}
        \includegraphics[width=\linewidth]{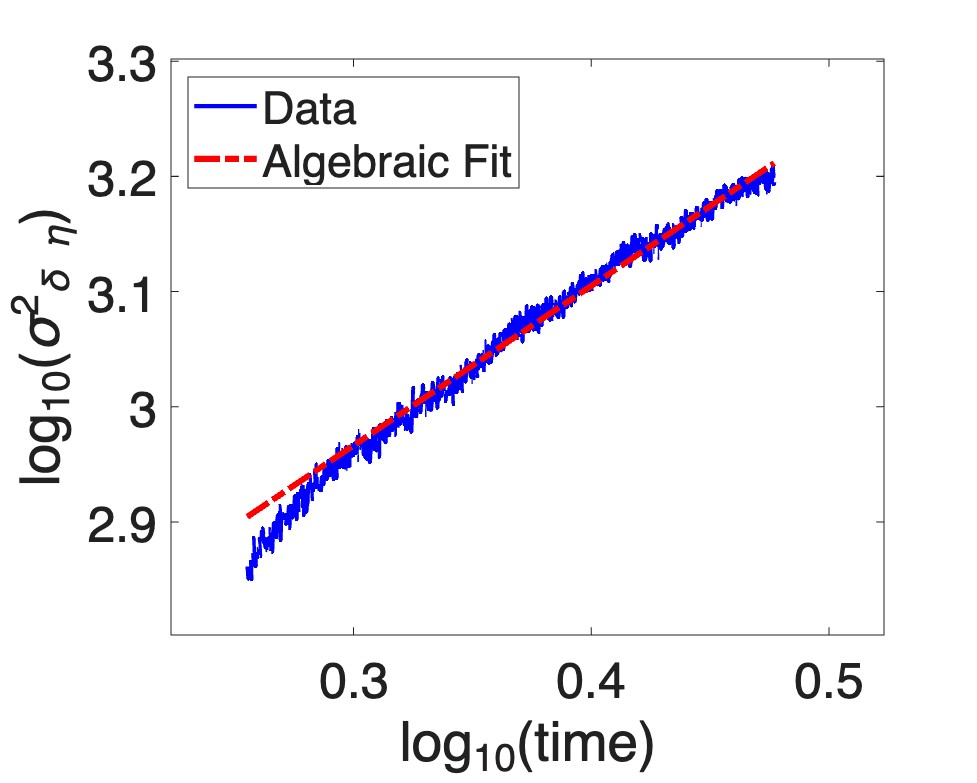}
        \caption{$K_{0}= 10-15 \ \text{MeV}$,$\sigma^2 \approx t^{1.4}$}
        \label{subfig2a}
    \end{subfigure}
    \hfill
       \begin{subfigure}{0.45\textwidth}
        \includegraphics[width=\linewidth]{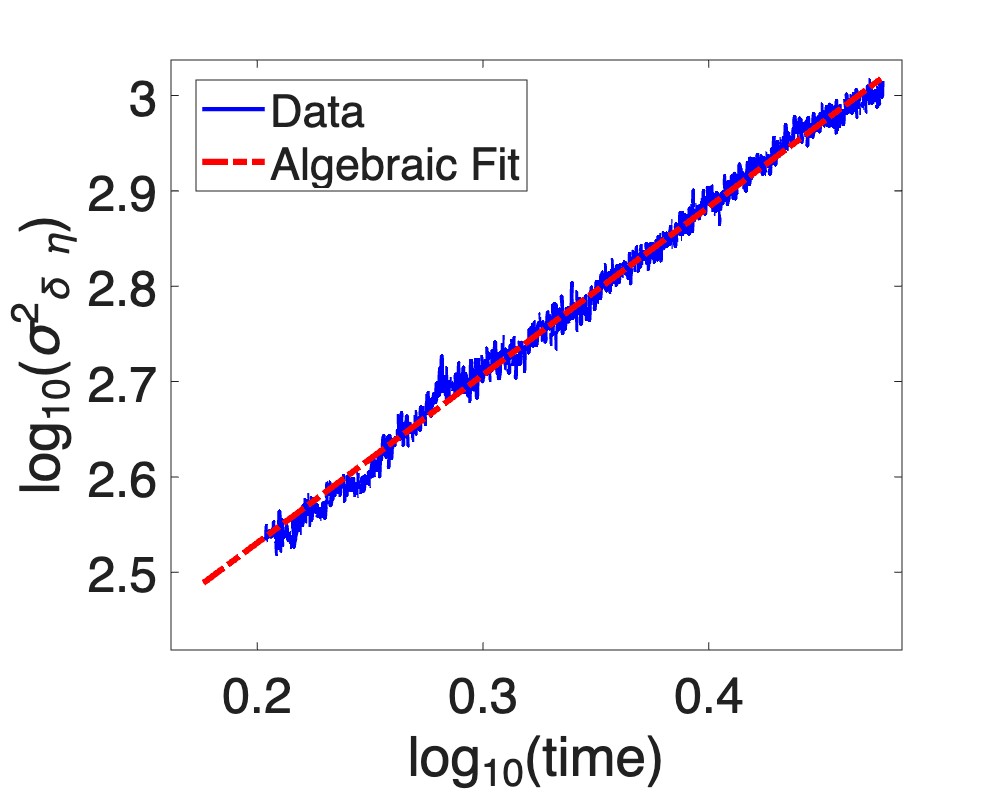}
        \caption{$K_{0}= 16-20\text{MeV}$,$\sigma^2 \approx t^{1.75}$}
        \label{subfig2b}
    \end{subfigure}
    \caption{ Log-log plots of the variance of $\delta \eta$ versus time with 2500 particles in each case. The blue curve represents the asymptotic behavior of $\log_{10}({\sigma^2}_{\delta \eta}(t))$, and the red dashed curve shows the linear fit used to determine the scaling of $\delta \eta$ variance with time.}
    \label{diff_scaling_vareta}
\end{figure*}

Figure \ref{diff_scaling_vareta} shows ${\sigma_{\delta \eta}}^2$ as function of $t$ for different initial relativistic kinetic energy $K_0$ intervals, where $ K = \frac{m_e c^2}{e}\left(\gamma - 1\right)$, and $\gamma = \left(1 - v^2/c^2\right)^{-1/2}$ is the relativistic Lorentz factor.
In the  $\log-\log$ scale we are using, the algebraic scaling of the variance manifests as the linear scaling $\log \left[{\sigma_{\delta \eta}}^2\right] \sim \alpha \log t$ with the scaling exponent $\alpha$ corresponding to the slope of the curves plotted in Fig. \ref{diff_scaling_vareta}.
% Figure  shows the time evolution of the mean pitch angle displacement for different initial kinetic energy ranges. A comparison of the subfigures indicates that for $KE_{initial}=1-10 \ \text{MeV}$, the mean pitch angle displacement evolves faster and reaches a higher value after $1m\text{s}$ than for $KE_{initial}>10\ \text{MeV}$. 
%The slope of these plots helps to identify the nature of the underlying transport process: if the y-axis in these plots have a linear relationship with the quantity on x-axis, i.e., $\log({\sigma^2}) \propto \alpha \ \log(t) \implies {\sigma^2} \approx t^{\alpha}$, the scaling exponent $\alpha=1$ indicates diffusion, $\alpha>1$ indicates super-diffusion, $\alpha<1$ indicates sub-diffusion \onlinecite{Diego2005}. 
It is observed that the pitch angle scattering of REs is sub-diffusive with scaling exponent $\alpha=0.75$ for REs with initial energies in the range $K_{0}\in [1,5] \ \text{MeV}$, slightly super-diffusive with $\alpha=1.2$ for $K_{0}= \in [5,10] \ \text{MeV}$,  and strongly  super-diffusive with $\alpha>1.4$ for $K_{0}>10 \ \text{MeV}$ .

\begin{figure}
\centering
\includegraphics[width=1.0\linewidth]{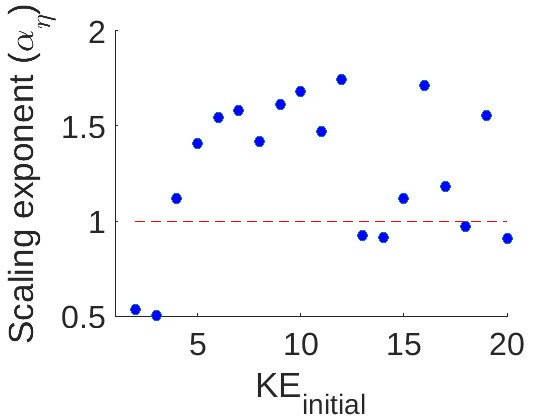}
\caption{Scatter plot of scaling parameter $\alpha$ depicted by blue dots (where, $\sigma^2 \propto t^\alpha$) for pitch angle displacements versus the initial kinetic energy of runaway electrons. The red dashed line represents the diffusive scaling corresponding to $\alpha=1$. Each marker on the horizontal-axis corresponds to a $1\ \text{MeV}$-wide energy bin of initial kinetic energy, where each $KE_{initial}\in[K_0,K_0+1\ \text{MeV}]$ for $K_0 =1, 2, \ldots, 19~\mathrm{MeV}$.}
\label{total_scaling_eta}
\end{figure}

To get further insights into the energy-dependent non-diffusive pitch angle displacements, we show in Fig.~\ref{total_scaling_eta} the value of the scaling exponent $\alpha$ for finer intervals of initial runaway electron energies. It is concluded that the RE pitch angle scattering exhibits diffusive behavior only for few initial kinetic energy ranges: $12-14$ MeV, $17-18$ MeV and $19-20$ MeV. In contrast, for the most initial energies, namely  $4-12\ \text{MeV}$ and $15-17\ \text{MeV}$, the pitch angle scattering is super-diffusive. Sub-diffusive behavior is observed at lower initial energies in the range of $1-3\ \text{MeV}$.
% 
%These results essentially highlight that the phenomenon of RE pitch angle scattering in the presence of whistler fields is non-diffusive in the sense that the transport scaling of RE pitch angles with time indicates diffusive, sub-diffusive, or super-diffusive transport depending upon the initial kinetic energy of the runaway electrons. 
These results challenge the assumption that the RE-whistler interaction is a diffusive phenomenon, and show that the RE transport in pitch angle space is typically non-diffusive and non-trivially dependent on REs kinetic energy. 
\par
We conclude this section discussing the implications of the previous results on the applicability of quasilinear diffusion (QLD) in the study of RE interacting with whistler fields.
Quasilinear diffusion theory provides a statistical description of how electromagnetic fluctuations produce diffusion in pitch-angle space by modeling the long-time cumulative effect of many weak wave–particle interactions in a plasma. The theory assumes that the fluctuating electromagnetic fields are small compared with the equilibrium fields, so particle orbits are only weakly perturbed and resonant interactions produce small incremental changes in pitch angle. Physically, particles are assumed to experience many small “kicks” rather than a few large scattering events. The theory also relies on the random-phase approximation, whereby the wave spectrum consists of many waves with uncorrelated or rapidly decorrelating phases. Under these assumptions, runaway electrons interacting with whistler waves are expected to exhibit ordinary pitch-angle diffusion, Gaussian-like transport statistics, and gradual energy redistribution.

The previously discussed emergence of Laplace distributions in the pitch-angle increments (see Figs.~\ref{PDF1} and \ref{PDF2}), in contrast to the Gaussian distributions expected from standard diffusion theory, calls into question the validity of the assumptions underlying quasilinear theory within the time-scales of the numerical study presented here. In general, stochastic processes characterized by Laplace statistics exhibit more frequent large deviations, stronger intermittency, and sharper localized events than Gaussian processes.
In the context of pitch-angle increments, the observed Laplace distributions suggest that the evolution of RE pitch angles does not appear to be governed solely by the accumulation of many small localized "kicks", as assumed in quasilinear theory. Instead, the dynamics appear to involve intermittent strong scattering events, coherent resonant interactions with whistler modes, and bursty transport in pitch-angle space for timescales of interest in the present study.
The non-diffusive scaling observed in Figs.~\ref{diff_scaling_vareta} and \ref{total_scaling_eta} provide further evidence that conventional quasilinear diffusion does not fully describe the pitch-angle dynamics studied here.

\section{Kinetic Energy Evolution}
\begin{figure*}[htb!]
    \centering
    \begin{subfigure}{0.45\textwidth}
        \includegraphics[width=\textwidth]{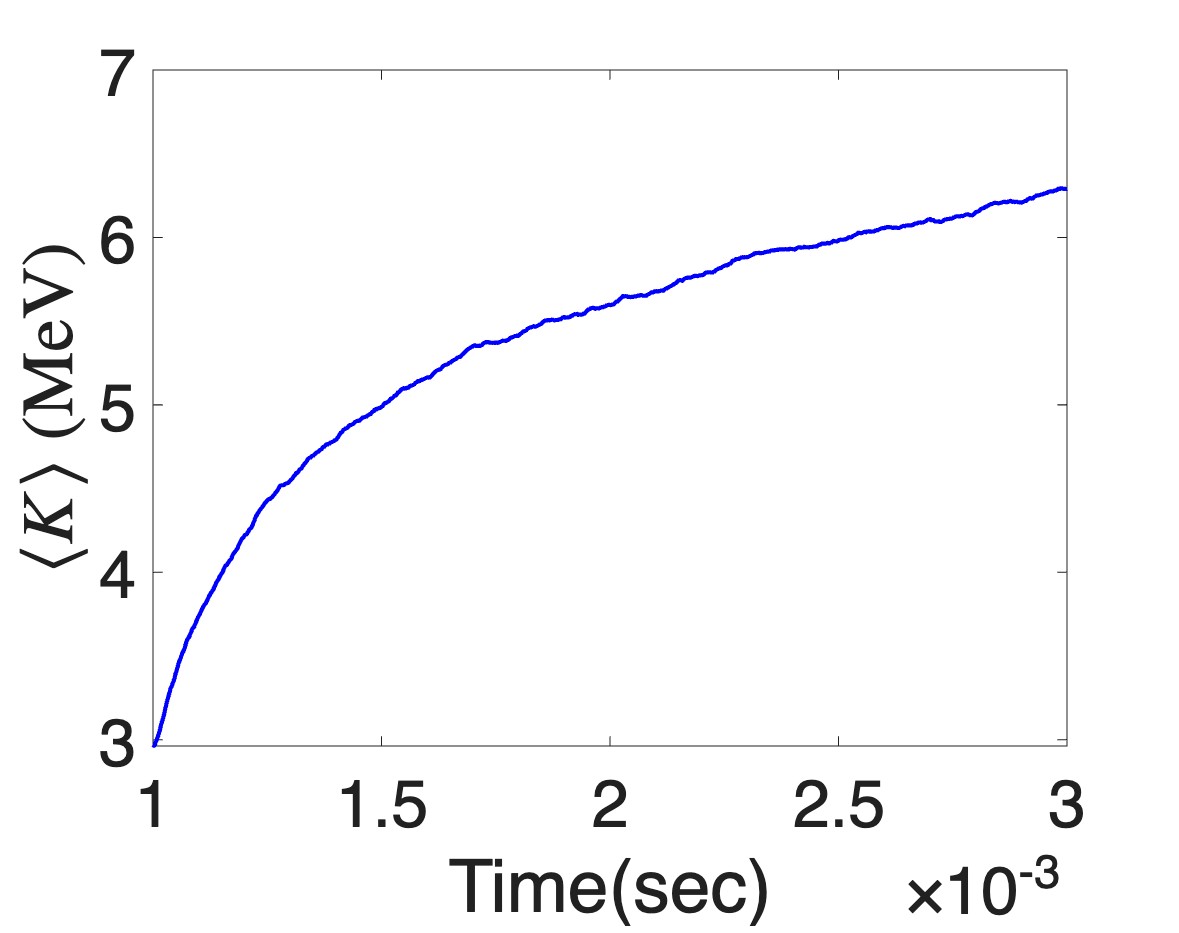}
        \caption{ $K_{0}= 1-5 \ \text{MeV}$}
    \end{subfigure}
    \hfill
    \begin{subfigure}{0.45\textwidth}
        \includegraphics[width=\textwidth]{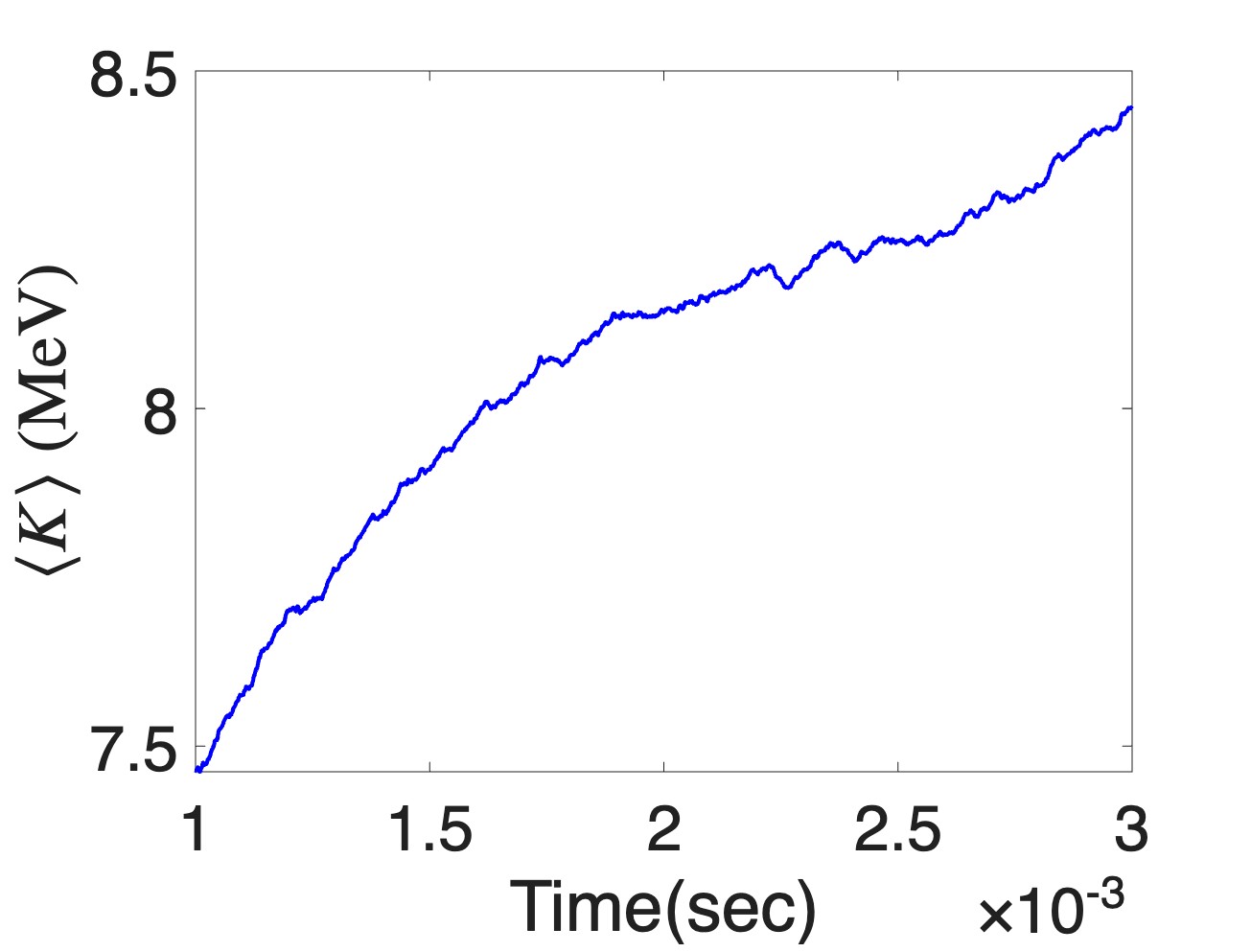}
    \end{subfigure}
  \\
    \begin{subfigure}{0.45\textwidth}
        \includegraphics[width=\textwidth]{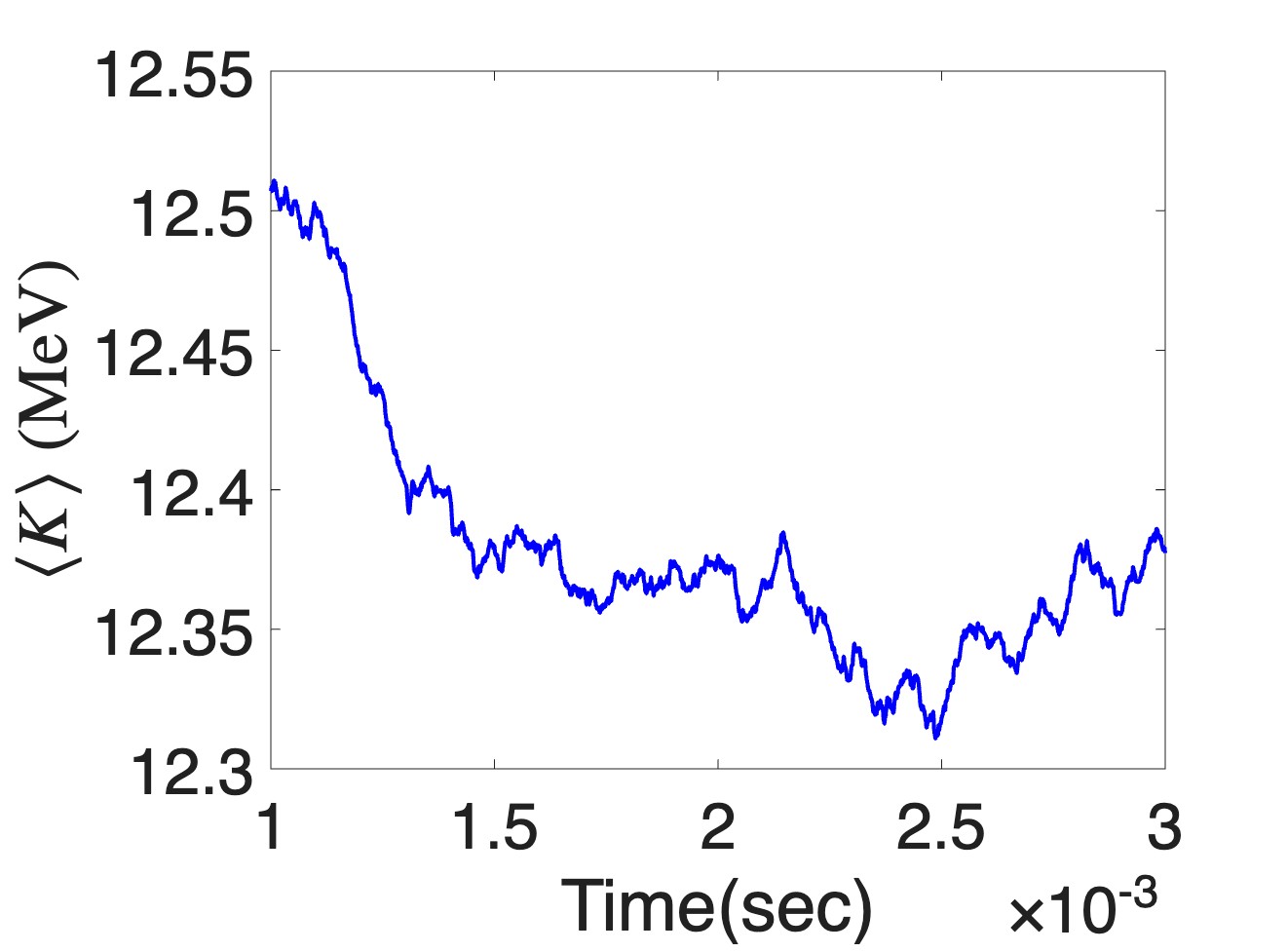}
        \caption{$K_{0}= 10-15 \ \text{MeV}$}
    \end{subfigure}
    \hfill
    \begin{subfigure}{0.45\textwidth}
        \includegraphics[width=\textwidth]{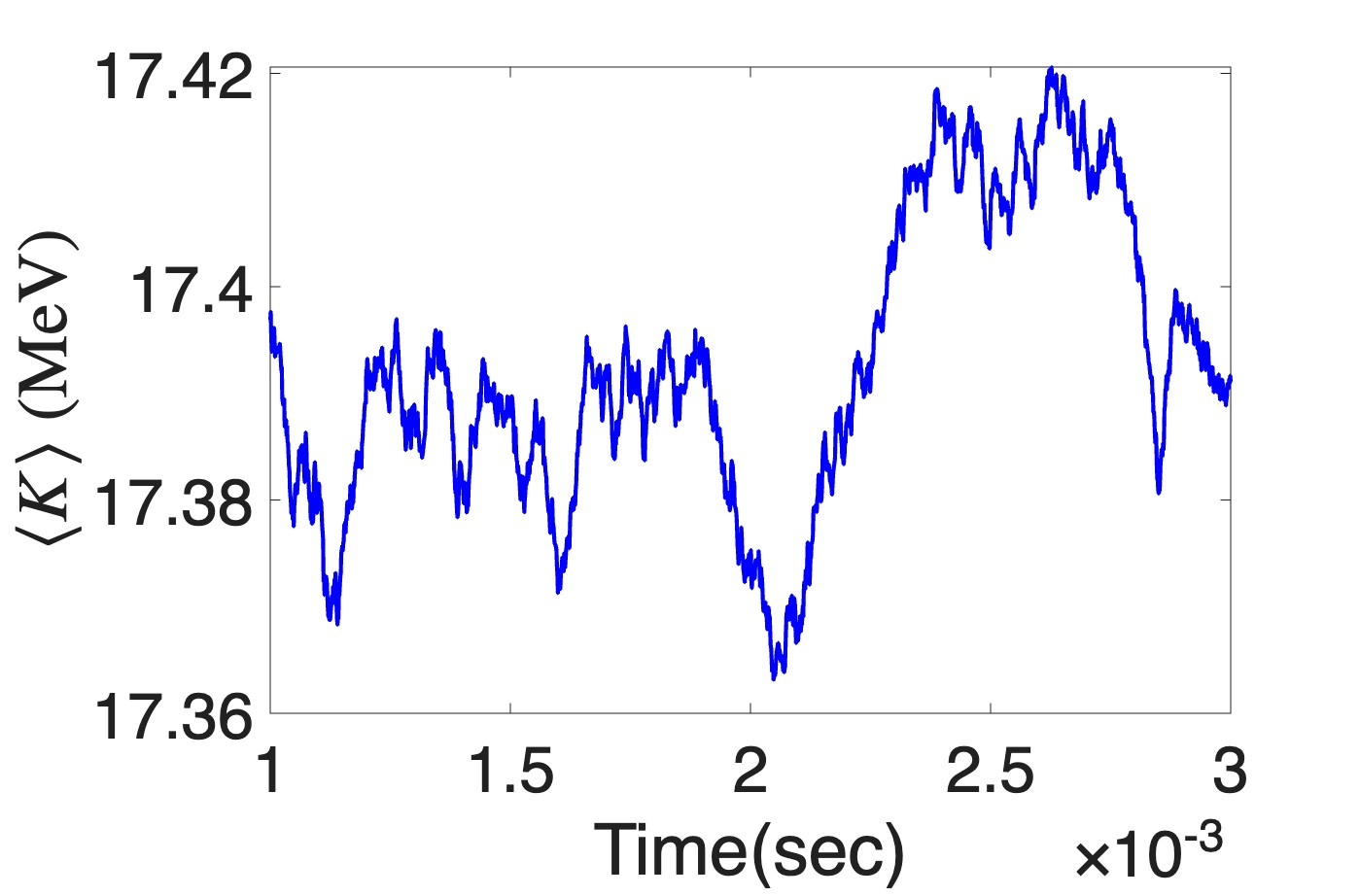}
        \caption{$K_{0}= 15-20 \ \text{MeV}$}
    \end{subfigure}
    \caption{Time evolution of the ensemble-averaged RE kinetic energy $\langle K \rangle$, for RE populations grouped according to their initial kinetic energy ranges. Each subplot corresponds to an ensemble containing $\approx 2500$ particles. The REs are evolved for 2 ms from a collisionlessly evolved mono-pitch distribution initialized at $10^\circ$ pitch angle. }
    \label{evo_E_avg}
\end{figure*}

To identify the impact of whistlers on the kinetic energy of REs, in Fig.~\ref{evo_E_avg}, we show the time evolution of the ensemble-averaged RE energy by dividing the RE population in four intervals based on their initial kinetic energy: (a)$K_{0}=1-5\ \text{MeV}$, (b)$K_{0}=5-10\ \text{MeV}$, (c)$K_{0}=10-15 \ \text{MeV}$, (d)$K_{0}=15-20 \ \text{MeV}$. It is observed that the REs with initial kinetic energy in the 1-5 MeV range experience highest energy gain, with their ensemble-averaged kinetic energy approximately doubling during the whistler interaction, while those with $K_{0} >10 \ \text{MeV}$ exhibit modest energy gain of approximately $1\ \text{MeV}$. In contrast, the RE population with $K_{0}=10-15\ \text{MeV}$ shows a net decrease in its ensemble-averaged kinetic energy with time, indicating a deceleration of this RE population in the presence of whistler fields. Such energy loss is consistent with the possibility of energy transfer from REs to whistler field in a fully-coupled physical system as observed in DIII-D experiments \cite{spong}. For the highest-energy RE population, $K_{0}=15-20 \ \text{MeV}$, the ensemble-averaged kinetic energy remains nearly constant, exhibiting only small fluctuations over the simulation interval. 
%This trend suggests a possible resonant energy exchange mechanism at play between REs and whistler waves.
 
\begin{figure*}[htb!]
    \centering
    \includegraphics[width=1.0\linewidth]{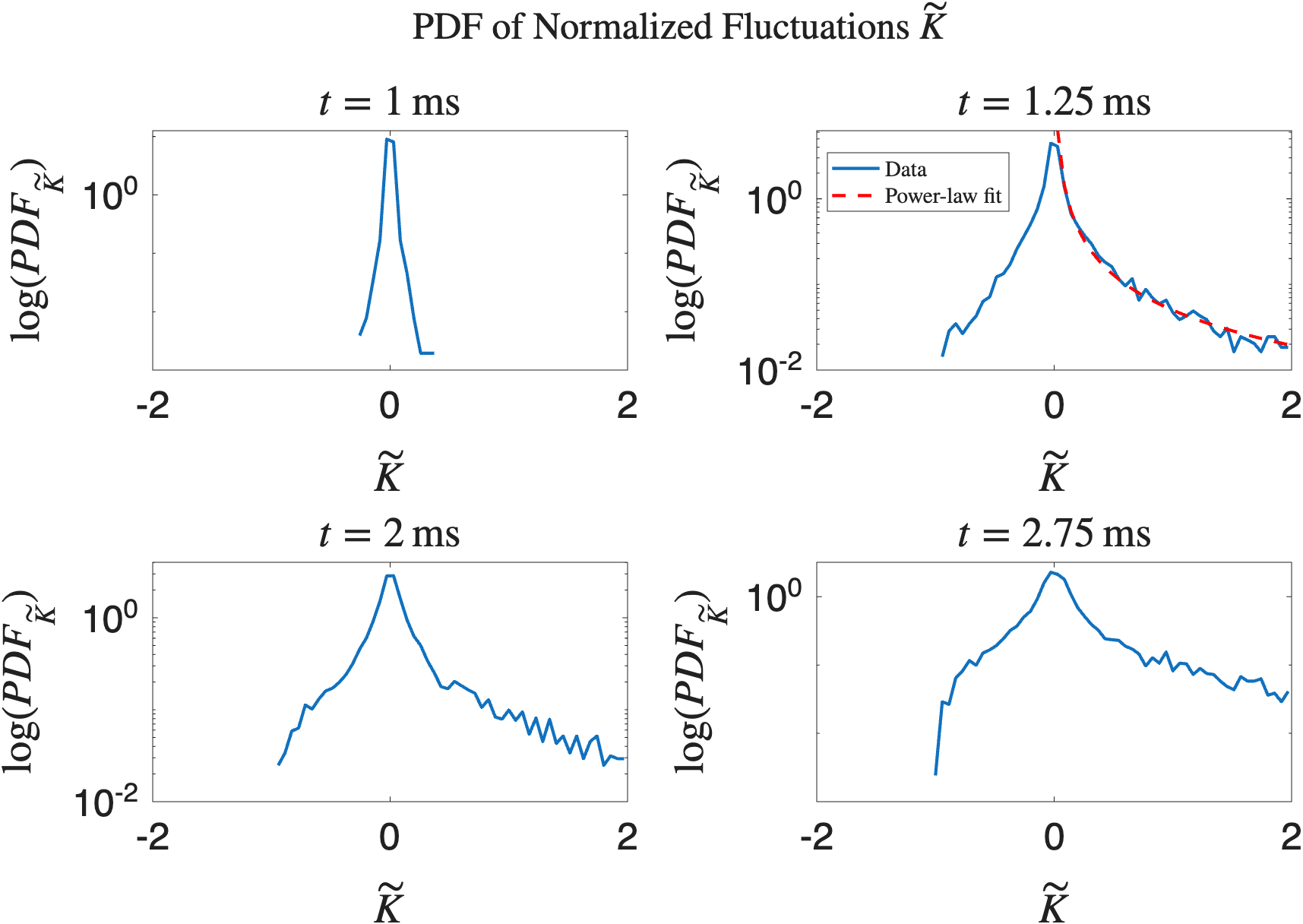}
    \caption{Probability distribution functions (log scale) of normalized kinetic energy displacements of REs at different simulation times.}
    \label{PDF_KE}
\end{figure*}

To further investigate this overall ensemble averaged increase in kinetic energy, in Figure \ref{PDF_KE} we show the log-scale plot of the PDFs of the normalized displacements of RE kinetic energy given by $\tilde{K}=(K_{t} -K_{0})/K_{0}$, where $K_t$ is the runaway electron kinetic energy at a given time $t$, $K_{0}$ is the initial kinetic energy, at different times in the simulation. It is clear from the histograms at different times that as the whistler fields were switched on, a fraction of REs gain and the remaining fraction loses their kinetic energy. However, the ensemble-averaged kinetic energy (as shown in Fig. \ref{evo_E_avg}) increases with time due to a larger fraction of runaways gaining energy as demonstrated by more positive displacements than negative in the normalized kinetic energy histograms after whistler fields are switched on. The positive tail of the normalized energy-kick distribution is well described by a power-law with exponent $\beta \approx 1.39$ as shown in Fig. \ref{PDF_KE} (sub-plot at $t=1.25 \ \text{ms}$). This behaviour deviates from the quasilinear diffusion expectations, this behavior indicates that the underlying wave–particle interactions are dominated by intermittent, rare events, in which a small fraction of particles experience large energy kicks. These events may contribute disproportionately to the increase in ensemble-averaged kinetic energy, particularly for runaway electrons with initial energies in the range $1-5 \ \text{MeV}$ as shown in Fig. \ref{evo_E_avg}. It is important to note here that currently there is no damping mechanism considered in the simulations. Hence, the gain or loss in RE energy is only attributed to the resonant acceleration or deceleration of REs in the presence of whistler fields and may lead to resonance broadening of REs with time, although a detailed study of resonance broadening is beyond the scope of this paper and may be addressed in a future publication.

%PDFs KE 
\section{Summary and conclusions}
  This work presents the first application of a coupled AORSA–KORC framework to study wave–particle interactions between runaway electrons (REs) and whistler waves in realistic tokamak geometry. Unlike previous approaches based on simplified quasilinear diffusion models or slab geometries, the present framework incorporates full-wave solutions of the Maxwell  system computed with AORSA and follows the evolution of RE full-orbit trajectories in these fields using the KORC code in a fully three-dimensional tokamak plasma.

A statistical analysis of the RE full-orbit trajectories computed with KORC in the presence of AORSA-generated wave fields reveals clear departures from Gaussian transport behavior. Whistler waves are found to induce significant pitch-angle scattering, quantified by increases in both the mean and standard deviation of the pitch-angle distribution of the RE ensemble. The skewness and kurtosis exhibit initial rapid growth, indicating the emergence of transient asymmetric and anomalously large pitch-angle excursions. Consistent with the behavior of the statistical moments, the probability distribution function (PDF) of the pitch angles broadens substantially and develops pronounced tails.

The PDFs of the pitch-angle increments, $\Delta \eta$, also exhibit strong deviations from Gaussian statistics and are well described by asymmetric Laplace distributions. The asymmetry of these distributions explains the increase in the skewness of the pitch-angle distribution, while their slowly decaying tails indicate the presence of intermittent large pitch-angle kicks, consistent with the observed increase in kurtosis.
The time evolution of the variance of pitch-angle displacements in the presence of whistler fields demonstrates anomalous, non-diffusive scaling that depends on the initial RE energy and can be either sub-diffusive or super-diffusive. 

Together, these results indicate significant departures from the statistical behavior expected from conventional quasilinear diffusion over the timescales considered here. In contrast to the quasilinear assumption that particles experience many small, symmetric pitch-angle kicks leading to Gaussian transport statistics, the present simulations reveal strongly intermittent and asymmetric large excursions in pitch-angle space. As a consequence, the pitch-angle distributions exhibit anomalously large skewness and kurtosis, together with non-Gaussian transport statistics and anomalous scaling of the variance. Departures from quasilinear dynamics are also observed in the energy evolution. In particular, the presence of slowly decaying power-law tails in the PDFs of normalized energy kicks indicates that wave–particle interactions are dominated by intermittent rare events in which a small fraction of particles experience disproportionately large energy changes.

The non-diffusive statistics identified in this work provides a basis for the development of reduced transport models capable of capturing non-diffusive pitch-angle scattering and intermittent wave--particle interactions in realistic tokamak geometry. These findings also motivate further investigation of RE mitigation through enhanced radiation damping, where externally launched waves could increase RE pitch angles and thereby enhance synchrotron radiation losses. It is important to emphasize that the statistical analysis presented here is based on finite-time numerical simulations, and the observed scaling may or may not represent the true asymptotic dynamics. Establishing the asymptotic regime would require simulations over substantially longer times and a systematic demonstration of convergence of the statistical scaling. Our conclusions should therefore be understood as characterizing the transport over the finite, but physically relevant and well-resolved, timescales considered in this study. 
\\
\clearpage
\begin{acknowledgements}
The authors thank David L. Green (CSIRO) for valuable discussions on AORSA code. The authors also thank William W. Heidbrink (General Atomics) for providing the DIII-D equilibrium used in this study. This work was supported by the U.S. Department of Energy, Office of Science, Office of Fusion Energy Sciences, under contracts no. DE-FG02-04ER54742 and DE-AC05-00OR22725, and through the MiRACL FIRE Collaborative. This research used resources of the National Energy Research Scientific Computing Center (NERSC), a Department of Energy User Facility (project m3236-2024)
\end{acknowledgements}
\section*{Data Availability Statement}
The data that supports the findings of this study are available from the corresponding author upon reasonable request.

\bibliographystyle{unsrt} 
\bibliography{aipsamp}
\end{document}